\documentclass[onecolumn,authoryear]{els-mrw} 
\usepackage{array}
\usepackage{multirow}
\usepackage{textcomp}
\usepackage{booktabs}
\usepackage{graphicx}
\usepackage{tabularx}
\usepackage{pdfpages}
\usepackage{tikz}
\usepackage{tablefootnote}
\usepackage{textcomp}
\usepackage{verbatim}
\usepackage{makecell}
\usepackage{enumitem}
\usepackage{comment}
\usepackage{subcaption}
\usepackage[T1]{fontenc}
\usepackage{lscape}
\usepackage[list=true]{subcaption}
\usepackage{hyperref}
\let\cite\citep
\usepackage{textcomp}
\usepackage{soul}
\usepackage{txfonts}
\usepackage{helvet}
\usepackage{longtable}
\usepackage[table]{xcolor}
\usepackage{rotating}
\usepackage{tikz}
\usepackage{pgf-pie}
\usetikzlibrary{shapes, arrows, positioning, fit, calc}
\usetikzlibrary{arrows.meta,trees,positioning,fit,backgrounds,shapes.geometric,calc}

\begin{document}

%\chapter{Assistive Technologies to support Ageing: Navigating Human-Robot Interaction Challenges in Private and Residential Aged Care Homes}\label{chap1}
\chapter[Navigating human–robot interaction challenges]
{
Navigating human-robot interaction challenges in private and residential aged care homes}
%Human–Robot Interaction for Home and Aged Care: Challenges, Vision Systems, and Robotic Solutions.

%\begin{frontmatter}

\author[1,2]{David Ahmedt-Aristizabal}
\author[1]{Mohammad Ali Armin}
\author[2]{Akila Pemasiri}
\author[2]{Harshala Gammulle}
\author[2]{Tharindu Fernando}
\author[3,4]{David Silvera-Tawil}
\author[2]{Simon Denman}

\address[1]{\orgname{CSIRO}, \orgdiv{Imaging and Computer Vision Group}, \orgaddress{Canberra, Australia}}
\address[2]{\orgname{Queensland University of Technology}, \orgdiv{School of Electrical Engineering and Robotics}, \orgdiv{Signal Processing, Artificial Intelligence and Vision Technologies (SAIVT Lab)}, \orgaddress{Brisbane, Australia}}
\address[3]{\orgname{CSIRO}, \orgdiv{Australian e-Health Research Centre}, \orgaddress{Sydney, Australia}}
\address[4]{\orgname{UNSW Medicine \& Health}, \orgdiv{International Centre for Future Health Systems}, \orgaddress{Sydney, Australia}}

\maketitle

\begin{abstract}
This chapter explores how robotic systems and ambient intelligence can enhance care for older adults in residential aged care (RAC) homes and private homes. Although these settings differ, we frame both as extreme and challenging environments for human–robot interaction (HRI), characterised by cluttered and dynamic layouts, diverse resident needs, and limited on-site technical support. We review embodied and non-embodied systems that assist with daily living, safety monitoring, and social engagement, highlighting hybrid deployments where robots complement fixed and wearable sensors.
To ensure these technologies are usable and effective, design considerations consider the cognitive, physical, and social needs of older adults. Interaction design prioritises accessibility for older adults with low digital literacy, using robust but simple voice interfaces, small sets of non-verbal cues, accessible screen-based companions, and safe object handovers. 
Privacy, trust, and acceptability are addressed via practical strategies such as on-device inference, data minimisation, visible consent controls, and transparent data handling aligned with emerging artificial intelligence (AI) and privacy regulation. 
Evaluation emphasises real-world metrics including false alarms per day, time to recover from faults, staff workload, and long-term system reliability. We close by outlining open research challenges and future directions for scalable, regulation-ready deployments that support aging with dignity, safety, and independence.
\paragraph{Key points:}
\begin{itemize}
    \item Characterises residential aged care and private homes as "extreme" HRI settings, with clutter, mobility aids, low digital literacy, and limited technical support.
\vspace{-6pt}
    \item Maps the ecosystem of assistive, service, social and telepresence robots working alongside ambient and wearable sensors in real care environments.
\vspace{-6pt}
    \item Describes core perception and learning technologies for safety, monitoring, and personalisation under privacy, compute, and data constraints.
\vspace{-14pt}
    \item Translates privacy, trust, and acceptability requirements into concrete interaction patterns and privacy-by-design safeguards.
\vspace{-6pt}
    \item Proposes evaluation approaches that go beyond accuracy, capturing workload, alarm burden, recovery from faults, and long-term engagement.
\vspace{-12pt}
    \item Identifies open research and deployment challenges for scalable, regulation-ready assistive robotics in RAC and private homes.
\end{itemize}
%
%\paragraph{Keywords:}
%Ageing in place; 
%Ambient intelligence; 
%Assistive robotics; 
%Edge and cloud computing; 
%Human–robot interaction(HRI); 
%Learning paradigms; 
%Older adults; 
%Privacy, trust and acceptability; 
%Residential aged care homes; 
%Scene understanding and mapping; 
%Sensor fusion; 
%Socially assistive robots; 
%Telepresence robots.

\begin{comment}
\begin{table}[ht]
\caption{Writing and review responsibilities by section}
\label{tab:writing-review-owners}
\centering
\begin{tabular}{@{}>{\raggedright\arraybackslash}p{0.55\linewidth}ll}
\toprule
\textbf{Section} & \textbf{Approx. status} & \textbf{With Who?} \\
\midrule
Abstract & 100\% & Finished \\
1.Introduction (1 page + 1 figure) & 100\% & Finished\\
2.Background, definitions, constraints environment (2 pages + 1 table) & 100\% &  Finished  \\
3.Interaction modalities and integration (2.5 page + 1 figure) & 100\% & Finished \\
4.1 Sensors and Devices (2.5 pages + 1 figure)  & 100\% & Finished \\
4.2 Scene understanding and mapping (0.5 page) & 100\% & Finished \\
4.3 Perceiving human's and their actions (0.5 page) & 100\% & Finished \\
4.4 Learning paradigms (3.5 pages + 1 figure) & 100\% & Finished \\
5.Privacy, Trust and Acceptability (2.5 pages + 1 figure) & 100\% & Finished \\
6.Benchmarking and Evaluation (2.5 pages + 1 figure) & 100\% & Finished \\
7.Open Research Challenges (1.5 pages) & 100\% & Finished \\
8.Future Research Directions (1 page) & 100\% & Finished \\
9.Conclusion & 100\% & Finished \\
\bottomrule
\end{tabular}
\end{table}
\end{comment}

\end{abstract}

%\end{frontmatter}

%-----------------------------------------------------------------------------------
%\input{sections/0_Ideas_structure}
%-----------------------------------------------------------------------------------

% --------------------------------------------------------------
\section{Introduction}

\begin{figure}[!t]
\centering
\includegraphics[width=0.8\linewidth]{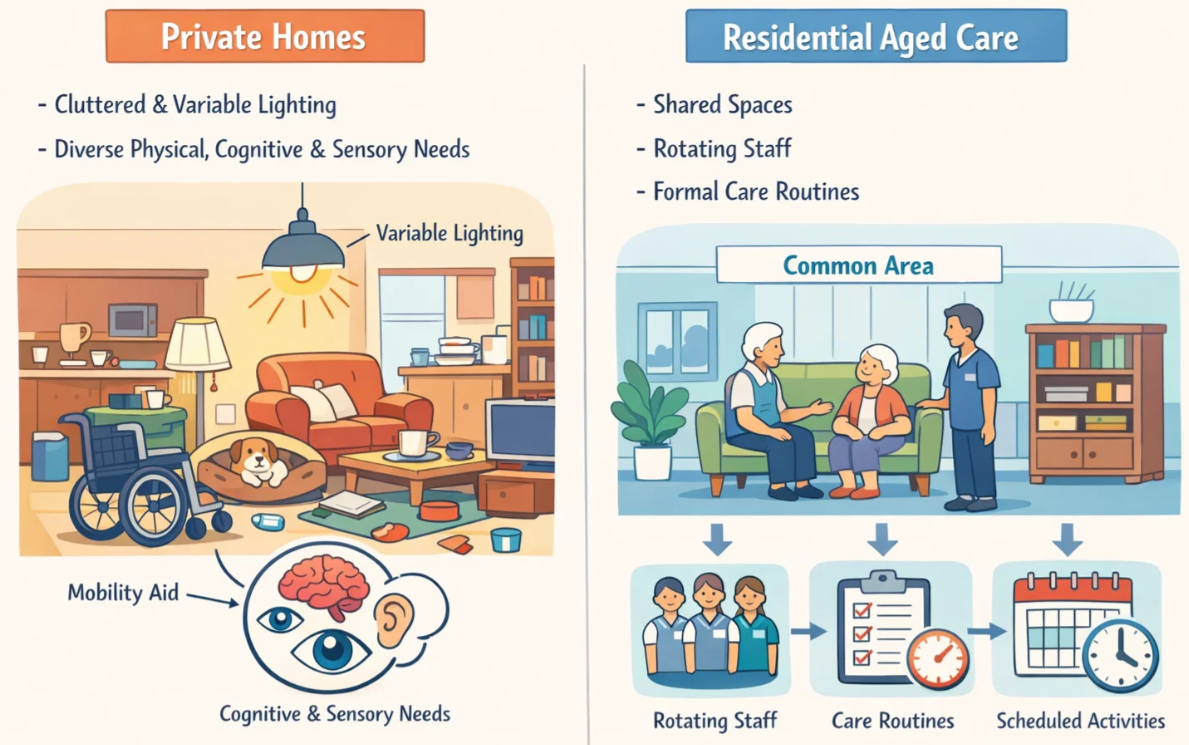}
\caption{
Example aged care environments illustrating clutter, mobility aids, and variable lighting that shape sensing and interaction: a private home living room (left) and a residential aged care corridor (right). 
}
\label{fig:env_examples}
\end{figure}

Human–robot interaction (HRI) in residential aged care (RAC) and private homes is gaining importance as populations age, and more people seek to remain in familiar environments while receiving safe, respectful support \cite{world2021decade}. At the same time, care providers face mounting pressures from staffing shortages and rising costs, while families seek care solutions that are dependable, dignified, and easy to use \cite{pu2019effectiveness}. Robots and artificial intelligence (AI) systems offer promising assistance for everyday tasks, social engagement, and safety monitoring, with the overarching goal of enhancing quality of life while preserving autonomy \cite{pu2019effectiveness,de2019would}. 
However, these environments present unique challenges for HRI. Private homes are highly heterogeneous, cluttered, and subject to varying lighting conditions, while residents often have diverse physical, cognitive, and sensory needs. RAC homes add further complexity through shared spaces, rotating staff, and formal care routines that must integrate with existing workflows~\cite{carros2022care,cavallo2018robotic,coradeschi2013giraffplus}. Taken together, these factors make HRI demanding, yet highly promising. We therefore consider RAC and private homes to be  \textbf{extreme HRI environments}, characterised by high day-to-day variability, limited on-site technical support, and stringent expectations around safety, privacy, and usability.
Fig.~\ref{fig:env_examples} illustrates typical RAC corridors and private living rooms, highlighting clutter, mobility aids, and variable lighting that shape sensing and interaction demands.

In such a context, privacy and safety are not optional. They are fundamental design constraints that determine what can be sensed, stored, and shared in daily life. Clear consent controls, data minimisation, and transparent system behaviour help build trust and support long-term acceptance. Equally critical are robust operation, fault tolerance, graceful degradation, and reliable recovery mechanisms that protect dignity and reduce caregiver burden \cite{world2024ethics,jin2022exploring}. These principles underpin the design and evaluation strategies explored throughout this chapter.

This chapter takes a primarily technical perspective on HRI to support older adults in both private and RAC homes. Our focus is on the sensors, perception algorithms, and robotic systems that enable safe and effective operation in these complex environments. While we acknowledge that human factors, usability, and organisational workflows are critical to success, the emphasis here is on the technical foundations that make these systems viable in practice, and we refer interested readers to complementary work on human factors, acceptability, co-design, and organisational implementation in care and home HRI~\cite{zhao2025social,jones2025acceptance,felber2024informing,sumner2021co,backaberg2024codesign}.

We review robotic platforms, vision-based perception, and ambient sensing systems designed for domestic and care-home contexts, examining how these technologies interact with people and identifying scenarios where embodied systems---such as physical robots---are essential for delivering meaningful support. Our analysis centres on practical challenges that constrain deployment, including robust perception in cluttered, lived-in spaces; interaction design for older adults with low digital literacy; and seamless integration with smarthome and care infrastructure. Alongside these technical considerations, we address cross-cutting factors such as safety, privacy, and trust, which are indispensable for sustained acceptance and long-term reliability.

%%%%%% Contributions and Chapter Roadmap
This chapter makes three main contributions. 
\begin{enumerate}
    \vspace{-6pt}
    \item It organises the design space for private and RAC homes into clear themes, including perception in lived-in spaces, interaction modalities tailored to diverse users, and deployment patterns that scale in practice.
    \vspace{-6pt}
    \item It maps privacy and trust requirements to concrete design strategies, translating ethical and regulatory expectations into actionable choices such as on-device inference, minimal data flows, visible consent and mute controls, and transparent data handling aligned with emerging AI and privacy regulations.
    \vspace{-6pt}
    \item It proposes practical evaluation considerations for real-world settings. These cover technical performance, safety, usability, and long-term reliability.
\end{enumerate}
\vspace{-6pt}
Together, these contributions provide a roadmap for designing and evaluating robots and ambient intelligence systems that are usable, trustworthy, and regulation-ready, enabling older adults to age with dignity, safety, and independence.

\vspace{2mm}

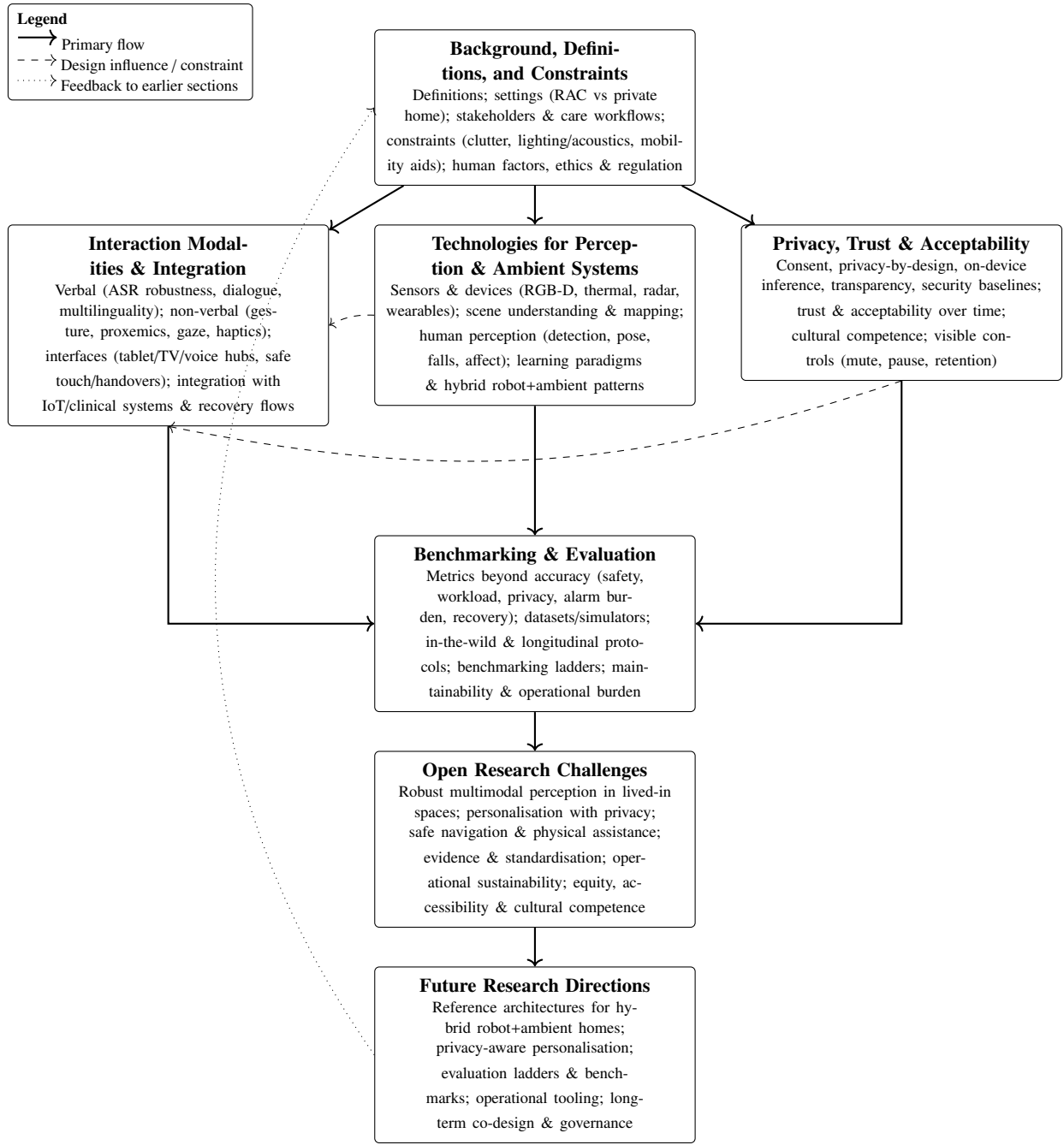
\begin{figure}[t]
\centering
\begin{tikzpicture}[
  node distance=6mm and 6mm,
  %node distance=10mm and 14mm,
  topic/.style={draw, rounded corners=2pt, align=center, inner sep=5pt, fill=white, text width=4.6cm},
  flow/.style={->, thick},
  influence/.style={->, dashed},
  feedback/.style={->, dotted}
]

% ---- Nodes ----
\node[topic] (background) {
\textbf{Background, Definitions, and Constraints}\\
\footnotesize Definitions; settings (RAC vs private home); stakeholders \& care workflows;\\
\footnotesize constraints (clutter, lighting/acoustics, mobility aids); human factors, ethics \& regulation
};

\node[topic] (interaction) [below left=of background, xshift=-1mm] {
\textbf{Interaction Modalities \& Integration}\\
\footnotesize Verbal (ASR robustness, dialogue, multilinguality); non-verbal (gesture, proxemics, gaze, haptics);\\
\footnotesize interfaces (tablet/TV/voice hubs, safe touch/handovers); integration with IoT/clinical systems \& recovery flows
};

\node[topic] (perception) [below=of background] {
\textbf{Technologies for Perception \& Ambient Systems}\\
\footnotesize Sensors \& devices (RGB-D, thermal, radar, wearables); scene understanding \& mapping;\\
\footnotesize human perception (detection, pose, falls, affect); learning paradigms \& hybrid robot+ambient patterns
};

\node[topic] (privacy) [below right=of background, xshift=1mm] {
\textbf{Privacy, Trust \& Acceptability}\\
\footnotesize Consent, privacy-by-design, on-device inference, transparency, security baselines;\\
\footnotesize trust \& acceptability over time; cultural competence; visible controls (mute, pause, retention)
};

\node[topic] (eval) [below=20mm of perception] {
\textbf{Benchmarking \& Evaluation}\\
\footnotesize Metrics beyond accuracy (safety, workload, privacy, alarm burden, recovery); datasets/simulators;\\
\footnotesize in-the-wild \& longitudinal protocols; benchmarking ladders; maintainability \& operational burden
};

\node[topic] (challenges) [below=of eval] {
\textbf{Open Research Challenges}\\
\footnotesize Robust multimodal perception in lived-in spaces; personalisation with privacy; safe navigation \& physical assistance;\\
\footnotesize evidence \& standardisation; operational sustainability; equity, accessibility \& cultural competence
};

\node[topic] (future) [below=of challenges] {
\textbf{Future Research Directions}\\
\footnotesize Reference architectures for hybrid robot+ambient homes; privacy-aware personalisation;\\
\footnotesize evaluation ladders \& benchmarks; operational tooling; long-term co-design \& governance
};

% ---- Primary flow (solid) ----
\draw[flow] (background) -- (interaction);
\draw[flow] (background) -- (perception);
\draw[flow] (background) -- (privacy);
\draw[flow] (interaction) |- (eval);
\draw[flow] (perception) -- (eval);
\draw[flow] (privacy) |- (eval);
\draw[flow] (eval) -- (challenges);
\draw[flow] (challenges) -- (future);

% ---- Cross-links (dashed) ----
\draw[influence] (perception.west) to[bend right=15] (interaction.east);
\draw[influence] (privacy.south) to[bend left=15] (interaction.south);

% ---- Feedback loop (dotted) ----
\draw[feedback] (future.west) to[bend left=35] (background.west);

% ----- Legend -----
\node[draw=black, rounded corners=2pt, align=left,
      inner sep=4pt, fill=white, font=\footnotesize,
      anchor=north west] at ([yshift=4mm]current bounding box.north west) (legend) {%
\textbf{Legend}\\
\raisebox{0.6ex}{\tikz{\draw[flow] (0,0)--(0.6,0);}} Primary flow\\
\raisebox{0.6ex}{\tikz{\draw[influence] (0,0)--(0.6,0);}} Design influence / constraint\\
\raisebox{0.6ex}{\tikz{\draw[feedback] (0,0)--(0.6,0);}} Feedback to earlier sections};

\end{tikzpicture}
\caption{Concept map of the chapter. Boxes summarise core sections; arrows show the primary reading flow, key cross-section design influences, and how future directions feed back into background assumptions as the field evolves.}
\label{fig:chapter-map}
\end{figure}

The remainder of this chapter is organised as follows. We begin by establishing the \textit{background, definitions, and constraints} of residential aged care and private home environments, clarifying system types, care settings, stakeholder roles, and the unique challenges these spaces pose for HRI. 
We then explore \textit{interaction modalities and integration}, including voice interfaces, simple non-verbal cues, and accessible screen-based companions. 
Next, we turn to \textit{technologies for perception and ambient sensing}, covering people and scene understanding, sensor choices, and edge and cloud design patterns, with a focus on hybrid robot and sensor configurations. 
This is followed by a discussion of \textit{privacy, trust, and acceptability}, translating ethical expectations into practical design strategies such as on-device processing, visible controls, and transparent data handling. 
We then address \textit{benchmarking and evaluation}, moving beyond accuracy to real-world implementation, including safety events, usability, and long-term reliability. 
Finally, we identify \textit{open research challenges}, including robust multi-modal fusion, privacy-preserving learning, and safe manipulation and operation in confined and cluttered spaces.
We then outline future research directions, proposing reference architectures, evaluation standards, deployment playbooks, and pathways to regulation-ready evidence. 
Fig.~\ref{fig:chapter-map} summarises these themes and how they interconnect across the chapter.

%------------------------------------------------------------------
%------------------------------------------------------------------

%---------------------------------------------
%---------------------------------------------
\section{Background, definitions, and constraints} 
\label{Sec:Background}

Before comparing specific technologies, it is essential to establish a shared understanding of the systems under discussion: what they are, where they operate, who interacts with them, and what successful deployment looks like. This section defines key system categories, situates them within RAC and private homes, clarifies stakeholder roles, and outlines the physical and human constraints that shape HRI in practice.

\subsection{Settings and deployment context}

We consider two primary settings: \textit{residential aged care (RAC) homes} and \textit{private homes} (ageing-in-place), as summarised in Table~\ref{tab:care-vs-home}. 
RAC homes offer structured environments with shared spaces, rotating staff, and formal procedures, enabling more managed deployments but requiring alignment with clinical governance, infection control, and the needs of multiple residents. In contrast, private homes are highly heterogeneous in layout, lighting, furnishings, and daily routines with minimal technical support and informal care arrangements. 

In both environments, tasks can be clinical, such as rehabilitation exercises or vital sign monitoring prompts; or non-clinical, including companionship, reminders, and safety checks. While clinical tasks require stricter documentation and oversight, non-clinical tasks are also essential for supporting well-being and daily routines. 
These tasks must demonstrate clear benefits to users and carers, while systems must remain low effort to operate and maintain~\cite{zieni2025interdisciplinary,boudouraki2025responsible,alshamsi2022robotics}. In addition, both settings impose stringent expectations around safety, privacy, and usability, making them extreme environments for HRI~\cite{carros2022care}.

\begin{table}[t]
\centering
\caption{Contrasting residential care homes and private homes as deployment contexts for HRI.}
\label{tab:care-vs-home}
\begin{tabular}{p{0.26\textwidth} p{0.32\textwidth} p{0.32\textwidth}}
\toprule
\textbf{Dimension} & \textbf{Residential care homes} & \textbf{Private homes} \\
\midrule
\textbf{Supervision and presence of staff} 
& Staff are usually on-site, with scheduled rounds and defined escalation pathways. Monitoring can be continuous in some areas and intermittent in others. 
& No formal on-site staff. Support is typically provided by family, visiting carers, or community services, often with long gaps between visits. \\
\addlinespace[0.5em]
\textbf{Policy and governance context} 
& Operate under aged care regulations, clinical governance, infection prevention policies, and facility-level procedures. Technology must fit accreditation and audit requirements. 
& Governed mainly by consumer protection, tenancy or home ownership arrangements, and general privacy and safety laws. Policies and record-keeping practices vary widely by household. \\
\addlinespace[0.5em]
\textbf{Environmental variability} 
& Layouts are somewhat standardised within a facility, with repeated room types and shared areas, although furniture and equipment may still move. 
& Highly heterogeneous and personalised. Floor plans, furniture, lighting, and clutter vary from home to home and change over time. \\
\addlinespace[0.5em]
\textbf{Available technical and care support} 
& Some access to internal information technology and maintenance teams, plus nursing and care staff who can help reset or reposition devices when trained. 
& Very limited technical support on site. Households rely on remote support, ad hoc local help, or do it themselves, often with low digital literacy. \\
\addlinespace[0.5em]
\textbf{Mobility aids and assistive equipment} 
& High prevalence of walkers, wheelchairs, hoists, oxygen stands, and shared medical equipment in corridors and rooms, with associated trip and collision risks. 
& Mobility aids may be present but are more idiosyncratic and less standardised. Distribution across homes is uneven, and storage locations can change frequently. \\
\bottomrule
\end{tabular}
\vspace{-6pt}
\end{table}

%-----
\subsection{Robotic and ambient systems considered}

We adopt the nomenclature introduced in \cite{SilveraTawil2024RoboticsSurvey}, and group assistive technologies into five categories. 
\textbf{Socially-assistive robots (SARs)} provide support primarily through social interaction rather than physical intervention. They aim to engage users emotionally and cognitively, offering companionship, therapeutic support, and motivation for tasks such as medication adherence, exercise, or learning. 
\textbf{Assistive robots} support individuals with physical disabilities in performing activities of daily living (ADLs), including mobility, personal hygiene, eating, and rehabilitation. This category includes robotic arms, smart wheelchairs, exoskeletons, and wearable devices that directly influence these movements and physical function. 
\textbf{Teleoperated robots} enable remote delivery of healthcare or support services. These systems are controlled by clinicians or carers at a distance and include telepresence platforms such as mobile video call devices that allow remote staff to navigate through a space and interact with residents. \textbf{Service robots} assist with logistical, repetitive, or physically demanding tasks such as delivering medications or disinfecting rooms. These robots typically have limited interactions with patients and pose minimal safety risks. \textbf{Ambient intelligence systems}, while not robots in a strict sense, include wearable and environmental sensors (e.g., cameras, microphones, radar, pressure mats) that monitor activities and detect risks in the background. In practice, these categories often overlap \cite{cruces2024socially,sandhu2024}. For example, a mobile robot may be paired with ambient sensors and a smart display to form an integrated system for reminders, safety monitoring, and family communication~\cite{yeo2023unpacking,carros2022care}.

%-----
\subsection{Stakeholders and roles}
HRI in these contexts involves multiple stakeholders with different priorities~\cite{de2019would,sharkey2012granny,boudouraki2025responsible,veling2024assistive}. 

\noindent\textbf{Older adults} are usually the intended beneficiaries, but not always the primary interface users. In both settings, older adults may interact directly with socially assistive robots, voice interfaces, and screen-based companions, while ambient sensors such as wearables and environmental devices operate unobtrusively in the background. In care facilities, residents may engage with robots during activities or therapy, but many functions---such as alert dashboards or sensor data interpretation---are mediated through staff.

\noindent\textbf{Care staff and clinicians} depend on these systems for actionable insights rather than raw data. Robots may deliver medication reminders or enable telepresence consultations, while ambient sensors provide fall alerts or routine monitoring. Integration with existing workflows is critical; alerts must be clear, reliable, and low-burden, and recovery from faults should not require technical expertise.

\noindent\textbf{Families and informal carers} often manage setup and troubleshooting in private homes. They value systems that are easy to install, require minimal maintenance, and provide reassurance without compromising privacy. Hybrid deployments---combining robots for social engagement with ambient sensors for safety---must balance functionality with transparency and control.
    
\noindent Finally, \textbf{facility managers, vendors, and regulators} shape deployment and governance. Managers prioritise reliability, infection control, and compliance. Vendors ensure interoperability, updates, and remote support. Regulators set privacy and safety baselines, influencing design choices such as on-device inference, visible sensing indicators, and consent mechanisms. Together, these roles underscore the need for technical solutions that align with human routines and regulatory obligations.

Due to different priorities in these groups, their interests can conflict. Clinicians may request granular data to support monitoring, while residents may prefer strong limits on what is recorded and shared. Staff may value automation that saves time, while families may prioritise human contact. To make such trade-offs explicit, we explore in later sections a simple task, risk, and burden lens that weighs outcomes against cognitive load for users and staff, as well as data exposure needs~\cite{yeo2023unpacking,carros2022care}.

%-----
\subsection{Physical and environmental constraints}
Both private and RAC homes present significant physical challenges for HRI, but the nature of these challenges differs across settings. Private homes are highly individualised and dynamic. Lighting conditions shift throughout the day, and the presence of pets, mobility aids, and visitors introduces unpredictability. These environments can feature uneven flooring, steps, tight spaces, cluttered surfaces, and narrow corridors, all of which complicate robot navigation and sensing. 
They can also be noisy, making speech recognition and sound-based alerts less reliable. Field studies consistently show that these everyday details shape how people use robots, often diverging from lab-based expectations~\cite{carros2022care,yeo2023unpacking}. 

RAC homes, while more standardised, involve shared spaces, equipment, infection control protocols, and frequent furniture changes. Robots and sensors must feature wipe-down surfaces, docking stations, and cable management solutions that do not obstruct movement or create new hazards~\cite{castelli2022review,ACSQHC_AgedCareIPC_2024}. Robots must cope with narrow corridors, occlusions, and strong lighting variations, while maintaining compliance with safety and hygiene requirements. Although care homes may offer more predictable layouts, they still involve frequent changes: assistive equipment is moved, chairs and tables are repositioned, and staff workflows vary across shifts. Pathways must also remain compliant with evacuation and manual handling requirements, which limit where larger mobile platforms can wait, turn, or dock between tasks.

Across both settings, robust perception, adaptive mapping, and conservative motion planning are essential. Sensor fusion, combining radar, vision, and audio, is often necessary to maintain reliability~\cite{yeo2023unpacking,piaggio2023use}. Ultimately, successful HRI relies on systems that are resilient to physical variability, respectful of user routines, and compatible with the practical realities of home and facility care.

%-----
\subsection{Human factors} 
Older adults often face barriers such as limited digital literacy, sensory impairment (e.g., hearing or vision), and varying degrees of cognitive decline. In private homes, these challenges are compounded by the absence of on-site technical support~\cite{zhao2025social}. Therefore, systems must be intuitive, require minimal setup, and degrade gracefully under faults. Interaction should favour simple, repeatable cues over complex dialogues, and sensing behaviour must be transparent to maintain trust. Staff need predictable alerts that do not add cognitive load or disrupt workflows, or systems risk being disabled despite technical capability.
    
\noindent These conditions have direct design implications for HRI:
\begin{itemize}
\vspace{-6pt}
    \item Perception must adapt to real lived-in spaces, supporting robust tracking with partial views, variable lighting, and coping with tight, cluttered spaces.   
    \vspace{-6pt}
    \item Interaction should prioritise simplicity and consistency. Short prompts, large font sizes, and clear confirmation phrases often work better than complex dialogue structures.
    \vspace{-6pt}
    \item Cognitive and sensory load must be managed by limiting unnecessary prompts and notifications, ensuring that alerts are meaningful and actionable.
    \vspace{-6pt}
    \item Workflows should degrade gracefully. If network connectivity is lost, core local functions should continue to operate, and the system should clearly flag which services are delayed or unavailable. 
    \vspace{-6pt}
    \item Privacy should be a default, with blurred or abstracted views when high detail is unnecessary, summaries stored instead of raw data, and visible indicators showing what is sensed and who can access the data~\cite{carros2022care}.
    \vspace{-6pt}
    \item Personalisation should be transparent and reversible, for example, allowing adjustments such as approach distance or prompt timing while providing clear controls to reset, pause, or review adaptations~\cite{papadopoulos2020caresses,yeo2023unpacking}.
\end{itemize}

In RAC homes, configuration and logging tools should align with existing roles and documentation practices so care workers can understand and adjust behaviour without specialist support. In private homes, setup and control should be simple enough for older people, family carers, and installers to manage with minimal ongoing effort. These principles align with universal design and accessibility standards such as WCAG 2.2~\cite{W3CWCAG22}.

%-----
\subsection{Ethical, legal, and cultural considerations}
Designing HRI for residential aged care and private homes is not only a technical challenge but also an ethical and regulatory one. Systems must comply with privacy, safety, and accessibility standards while operating in highly personal environments. Principles such as data minimisation, transparency, and informed consent are foundational, ensuring that sensing and decision-making respect user autonomy. These principles extend beyond interface design to system architecture, influencing choices such as on-device inference, short retention periods, and clear indicators of active sensing. Recent guidance for AI-enabled technology in health and disability settings expands on practical principles that align with HRI in care contexts~\cite{world2024ethics,Silvera-TawilAIFramework}. 

Compliance with emerging AI governance frameworks is increasingly critical. Regulatory and standards baselines are converging on risk-based governance. Teams can draw on the EU AI Act for duties applied to higher-risk systems, such as transparency, human oversight, post-market monitoring, and on ISO and IEC guidance, such as ISO and IEC 23894:2023 for risk management through the lifecycle of AI systems \cite{euaiact2024,iso23894}. In Australia, the Office of the Australian Information Commissioner’s Australian Privacy Principles sets duties on collection, consent, access, and security, including for audio and video from homes; and offers practical guidance for informed consent, retention limits, and access logs~\cite{oaicAPPs2022}. Residential care homes may also trigger additional obligations under health, aged care, or medical device regulations, whereas in private homes, many deployments are currently framed as consumer or lifestyle technologies even when they influence care, which creates a grey zone that designers should treat cautiously.

Beyond legal compliance, cultural competence and fairness are essential for inclusivity. Homes are personal spaces. People differ in language, routines, family roles, and how they read touch, direct their gaze, and the distance at which they interact with others. Trials of culturally competent social robots show that adapting greetings, topics, and forms of address can improve comfort and engagement, but only when changes are transparent and not deceptive~\cite{papadopoulos2020caresses}. Over-personalisation, particularly if it is perceived as manipulative or uncanny, can reduce trust and acceptance rather than improve it.

%-----------------------------------------------------------------------------------

%\clearpage
% --------------------------------------------------------------
\section{Interaction modalities and integration}
\label{Sec:Interaction}

\begin{figure}[!h]
\centering
\includegraphics[width=0.7\linewidth]{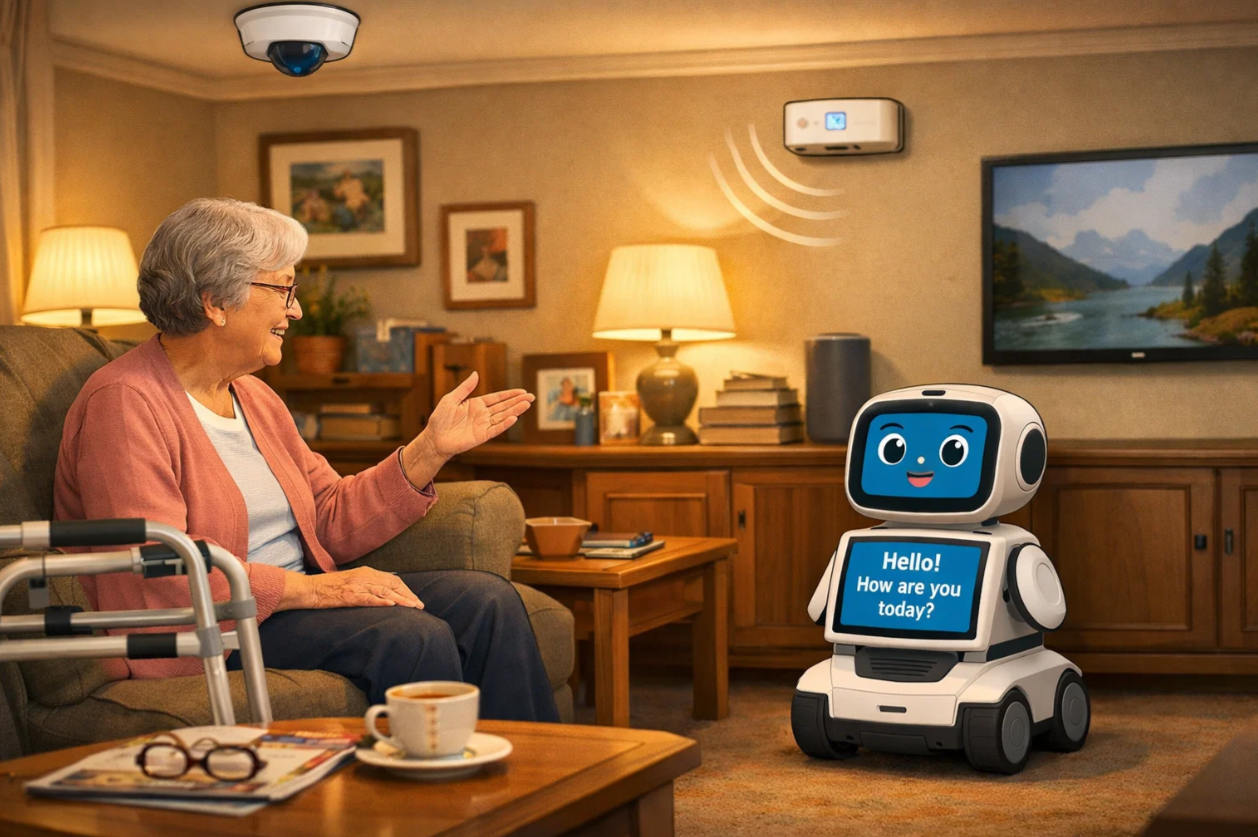}
\caption{
Example interaction scenario in a residential aged care lounge, showing a mobile social robot engaging an older adult via speech and on-screen prompts, alongside ambient devices such as a smart speaker, wall-mounted tablet, television, and ceiling and wall sensors. The scene illustrates how multimodal interaction, distributed sensing, clutter, mobility aids, and uneven lighting jointly shape design choices.
}
\label{fig:interaction_overview}
\end{figure}

In this section, we focus on how factors such as cluttered layouts, variable lighting, and the residents' diverse sensory and cognitive abilities shape interaction design for robots in RAC and private home care. We examine verbal and non-verbal modalities, fallback mechanisms, and integration patterns that aim to ensure reliability and accessibility in real-world conditions.
Fig.~\ref{fig:interaction_overview} illustrates a typical interaction scenario in a residential aged care lounge, with a social robot embedded within a wider ambient sensing infrastructure and operating under these real-world constraints.

\subsection{Interaction with robotic systems}

\subsubsection{Verbal interaction}
Verbal interfaces offer an intuitive way for older adults to interact with digital systems, particularly when hands are occupied or mobility is limited. However, these interfaces must be designed for the realities of home and RAC environments, where background noise, overlapping conversations, and age-related changes in speech are common. Far-field microphones often struggle with television or radio crosstalk, and multi-party dialogue can further degrade recognition accuracy. 
Large multilingual models trained on diverse data (e.g., Whisper and SeamlessM4T v2) have improved robustness to noise and support accented or code-mixed speech~\cite{radford2023robust,barrault2023seamlessm4t}. 

Despite these advances, recent benchmarks show that Automatic Speech Recognition (ASR) systems continue to underperform with older adult speech in noisy home settings, especially when speech is monotone or hesitant~\cite{escolano2025asr}. Field studies confirm these limitations, reporting lower accuracy rates and occasional delays. To mitigate these risks, designs should incorporate confirmation steps for safety-critical actions, provide fallback options such as a tablet, and favour short unambiguous utterances such as yes, no, or stop~\cite{cao2024factors,wang2023hyper,yeo2023unpacking}. Reliability improves when systems combine on-device voice-activity detection with basic beamforming. Providing an optional press-to-talk or hold-to-listen mode for critical intents, and applying diarisation and barge-in suppression for robots to ignore irrelevant audio sources, further enhances performance~\cite{radford2023robust,barrault2023seamlessm4t}. 
    
Interfaces should make microphone states visible (e.g., listening or muted), request explicit confirmations for irreversible actions such as calling an ambulance, and, where possible, perform on-device ASR while transmitting only transcripts or event summaries. 
Accessibility defaults remain essential, including large targets, high-contrast captions, and clear error messages~\cite{W3CWCAG22}. 
Interfaces should announce state changes (e.g., microphone muted or unmuted) for users with low vision. Finally, every interaction flow should include an error recovery ladder to maintain usability under failure conditions: Plan B, a fallback to a screen prompt, and Plan R, a retry step before escalation to human support.

\subsubsection{Non-verbal interaction}
Non-verbal cues, such as gestures, gaze, and posture, play an important role in making robots intuitive and socially acceptable. These modalities complement speech and provide alternative channels of communication for users with hearing impairments or limited verbal ability. However, their design must account for the sensory and cognitive diversity of older adults and the physical constraints of cluttered environments, as subtle cues can be  missed~\cite{morillo2023gaze,morillo2024age,urakami2023nonverbal}. 

Effective HRI in care settings favours a small and consistent, for example, point, wave, and open-hand stop, with slow, exaggerated motion profiles that remain legible under glare, clutter, and low light. Eye gaze and head orientation can support turn-taking and guidance, but should be complemented by redundant channels such as expressive lighting bands and large on-screen icons. These backups become primary when gesture confidence drops due to occlusion or poor lighting, with verbal confirmation where needed.

Physical interaction requires additional safeguards. Approach distance should be personalised rather than fixed, and interactions should remain natural to support engagement. For physical contact, such as handovers, robots could announce their intent verbally (e.g., "I’m passing the cup") and implement retreat behaviours when unexpected contact is detected~\cite{duan2024human}. It is important to keep interactions natural to improve engagement and enjoyment.

In cluttered rooms, gesture recognition should rely on short temporal consistency (around one second of a stable pose) and conservative confidence thresholds to avoid false positives. When gestures fail, expressive lighting and on-screen prompts should remain legible for users with low vision or higher cognitive load~\cite{urakami2023nonverbal,morillo2024age}. 

Cultural factors also influence non-verbal interaction. Cross-cultural studies show that gesture preferences vary between cultures, especially for greetings and abstract commands~\cite{shen2025ted}. For hygiene and safety, robots should use smooth, wipeable covers. Charging and docking stations should be placed away from spill zones, and from paths used by walking aids to reduce incidents \cite{ACSQHC_AgedCareIPC_2024}.

%-----
\subsection{Integration with the home and care ecosystem}
\label{integration_with_homecare_ecosystem}

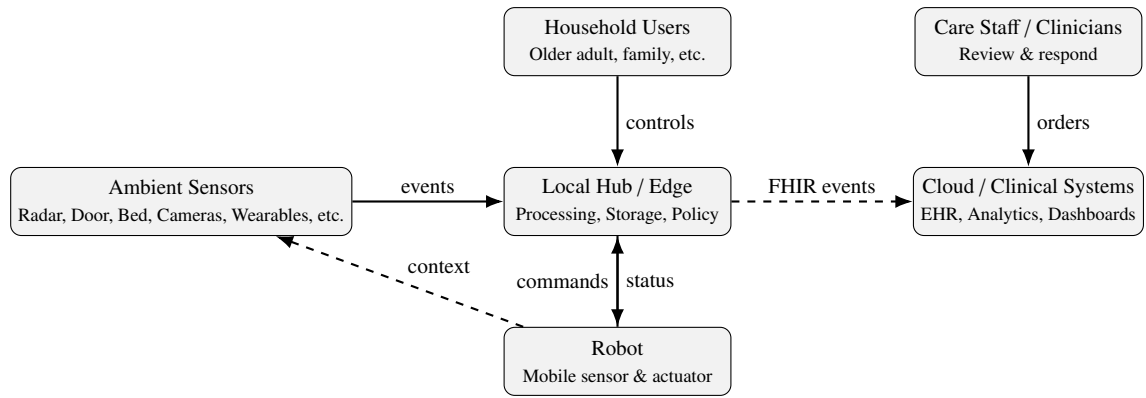
\begin{figure}[t]
\centering
\begin{tikzpicture}[
  font=\small,
  node distance=10mm and 16mm,
  box/.style={rectangle, rounded corners, draw=black, fill=gray!10, align=center, minimum height=9mm, minimum width=30mm},
  circlebox/.style={circle, draw=black, fill=gray!10, align=center, minimum size=9mm},
  line/.style={-Latex, thick},
  dashedline/.style={-Latex, thick, dashed}
]

% --- Left: Ambient sensors cluster ---
\node[box, minimum width=36mm] (ambient) {Ambient Sensors\\ \footnotesize Radar, Door, Bed, Cameras, Wearables, etc.};

% --- Center: Local hub / edge ---
\node[box, right=20mm of ambient] (hub) {Local Hub / Edge\\ \footnotesize Processing, Storage, Policy};

% --- Robot ---
\node[box, below=12mm of hub] (robot) {Robot\\ \footnotesize Mobile sensor \& actuator};

% --- Cloud / clinical systems ---
\node[box, right=24mm of hub] (cloud) {Cloud / Clinical Systems\\ \footnotesize EHR, Analytics, Dashboards};

% --- Household users ---
\node[box, above=12mm of hub] (hh) {Household Users\\ \footnotesize Older adult, family, etc.};

% --- Care staff / clinicians ---
\node[box, above=12mm of cloud] (clinicians) {Care Staff / Clinicians\\ \footnotesize Review \& respond};

% --- Flows ---
\draw[line] (ambient) -- node[midway, above]{events} (hub);
\draw[line] (robot) -- node[midway, right]{status } (hub);
\draw[dashedline] (hub) -- node[midway, above]{FHIR events} (cloud);

\draw[line] (hh) -- node[midway, right]{controls} (hub);
\draw[line] (clinicians) -- node[midway, right]{orders} (cloud);

% --- Optional direct links ---
\draw[dashedline] (robot) -- node[midway, above right]{context} (ambient);
\draw[line] (hub) -- node[midway, left]{commands} (robot);

\end{tikzpicture}
\caption{System architecture: robot and ambient sensors feed a local hub for edge processing and policy enforcement, while clinical summaries flow to cloud/Electronic Health Record (EHR) systems via standard interfaces built on Fast Healthcare Interoperability Resources (FHIR).}
\label{fig:system-architecture}
\vspace{-6pt}
\end{figure}
% -----------------------

Integration in home and care environments must address multi-actor workflows and limited on-site technical support. Systems should minimise setup burden and align with existing clinical and care data pathways to avoid creating isolated technology silos. Robots rarely operate alone; they function as part of a broader ecosystem that includes ambient sensors, local hubs, and cloud services. Fig.~\ref{fig:system-architecture} illustrates this architecture, where robots interact with IoT devices and clinical systems to enable edge processing, enforce privacy policies, and support data exchange.
    
Smart-home and IoT sensors, such as radar, wearables, and door or bed sensors, provide valuable context signals, including occupancy, routines, and sleep patterns. Robots can consume these signals while acting as mobile sensors and actuators, bridging gaps between fixed infrastructure and dynamic user needs. This hybrid approach improves robustness and enables richer interaction, for example, by combining fall detection from ambient sensors with verbal prompts or physical assistance from a robot. 

When clinical data must flow (e.g., fall alerts or adherence summaries to a clinician), standards-based integration is essential. Interfaces built on Fast Healthcare Interoperability Resources (FHIR) ensure events are recorded in existing electronic health record systems rather than bespoke pipelines~\cite{pavao2023fast,li2024review}. A practical pattern is to publish minimal event payloads (e.g., an \texttt{Observation} or \texttt{Detected Issue} with timestamp, severity, and a short rationale), maintain an audit log of ``who saw what and when'', and default to edge processing with only upstream summaries. Raw audio or video should be shared only under explicit consent and with clear retention limits, as elaborated in Section \ref{Sec:Privacy}.
Edge AI enables real-time decision-making but operates under constraints of compute, memory, and battery life~\cite{tahir2025edge}. Early deployments benefit from limiting the integration surface to a small set of well-defined events (e.g., possible fall, missed medication prompt, and failed dock), each with severity levels and human-readable rationales. This reduces maintenance and accelerates clinical sign-off. Aligning with national interoperability plans (e.g., Australia’s, 2023 to 2028 roadmap) and provider incentives for remote monitoring will further reduce deployment friction and support scalable adoption~\cite{adha2024interoperabilityprogress}.

%-----
\subsection{Design patterns and deployment insights}

\begin{figure}[!t]
\centering
\includegraphics[width=0.8\linewidth]{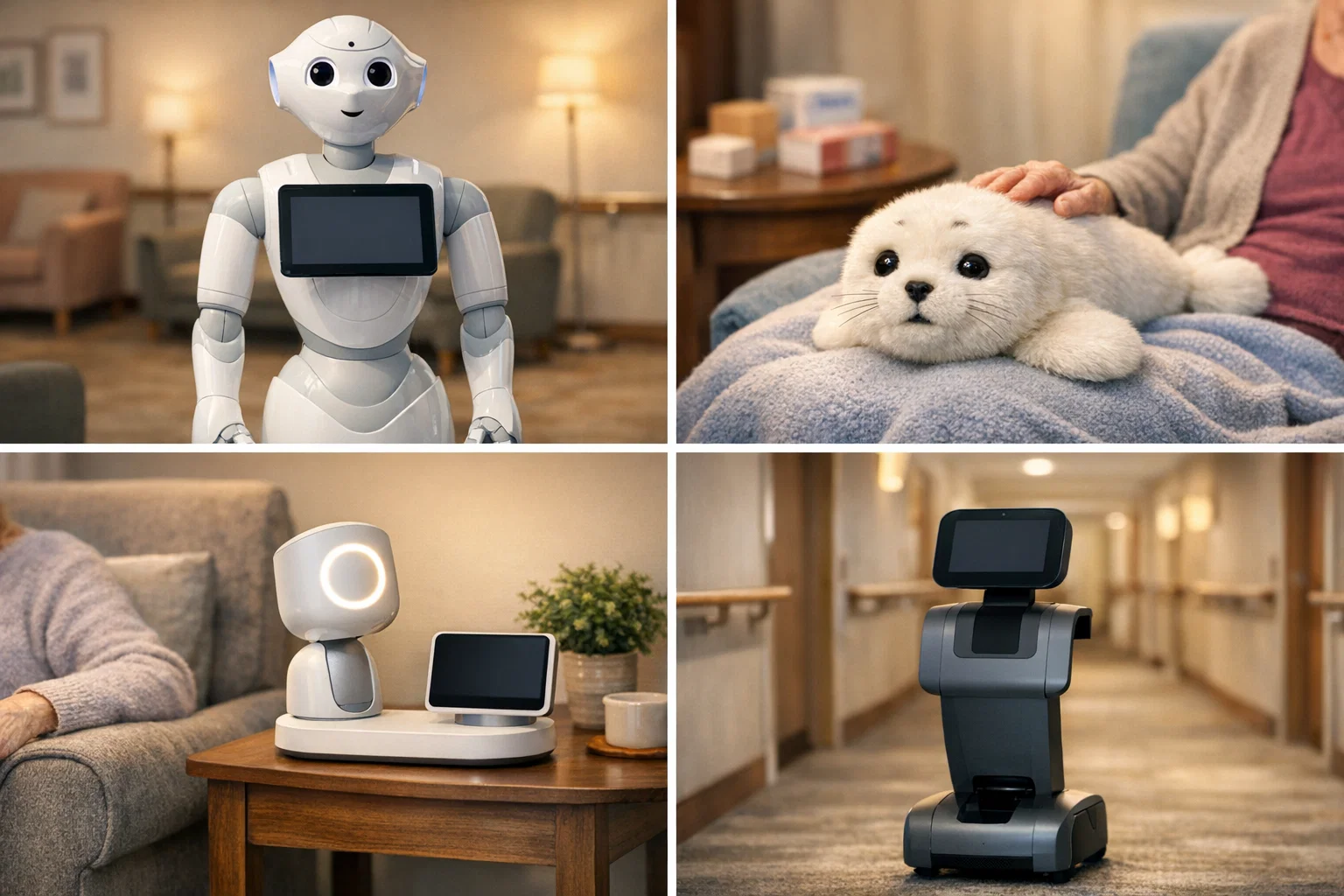}
\caption{
Examples of robot form factors used in residential aged care and private homes. 
Top left: humanoid service robot similar to Pepper. 
Top right: pet-like therapeutic robot similar to PARO. 
Bottom left: tabletop social companion similar to ElliQ. Bottom right: mobile telepresence/assistant robot similar to Temi.
}
\label{fig:rac_robots_collage}
\end{figure}

Field deployments in private and RAC homes consistently reveal patterns that improve reliability and acceptance. Across recent studies, four principles stand out: 
(i) Schedule interactions rather than relying on ad hoc use, as structured routines improve engagement and predictability; 
(ii) provide fast caregiver controls for setup and content customisation, reducing dependence on technical staff; 
(iii) enable fallback mechanisms so speech failures revert to large-screen prompts or alternative modalities; and 
(iv) instrument operations, including docking success, false speech activations, and alert rates to enable iterative improvement. These patterns create a direct bridge to the evaluation priorities in Section~\ref{Sec:Benchmark}, where reliability, recovery, and workflow fit become measurable outcomes rather than informal observations.
Fig.~\ref{fig:rac_robots_collage} illustrates four representative robot platforms that recur across residential aged care and home deployments, spanning humanoid service robots, pet-like therapeutic devices, tabletop companions, and mobile telepresence or assistant robots.

\subsubsection{Mobile social robots in RAC homes.}
Deployments of social robots such as Pepper (in a UK RAC home) highlight the importance of reliability and ease of customisation over feature breadth~\cite{manavi2025making}. In a three-month in-situ study, Pepper was used regularly, around 1 h/day, mostly in one-to-one sessions. Staff success depended on how easily they could tailor activities and routines, with reliability and time pressure acting as barriers. Follow-up work showed that providing care professionals with a lightweight mobile application to configure content and behaviours across multiple facilities increased appropriation and reduced reliance on technical staff~\cite{manavi2025making}.

\subsubsection{Telepresence for connection and clinical touchpoints.} 
Telepresence robots are valued for maintaining social contact and enabling remote clinical interactions, provided navigation is smooth and controls remain simple~\cite{yeo2023unpacking}. A 2024 qualitative study with frontline staff in hospitals and long-term care homes emphasised workflow fit, training, and clear escalation procedures as requirements for routine use in dementia care \cite{ren2024working}. Smaller trials indicate that repeated exposure improves comfort and perceived usefulness, although in-person contact remains important \cite{arthanat2024perspectives}. In practice, pairing autonomous navigation with scheduled sessions and a short menu of common tasks (e.g., check-in, family call, and medication prompt) reduces cognitive load for staff and residents.

\subsubsection{Companion/therapeutic robots.} 
Meta-analytic evidence suggests that PARO, the robotic seal, can reduce agitation and improve mood in dementia care when sessions are structured and integrated into activity schedules rather than left as ad-hoc interactions \cite{rashid2023effectiveness}. Recent clinical studies confirm sustained benefits for neuropsychiatric and social outcomes when programs are maintained and staff are trained~\cite{nam2025effectiveness,che2025effectiveness}. These findings underscore that companion devices should be treated as part of a program with a defined dose and frequency and facilitator training, rather than simply being a device left in a common room.

\subsubsection{Home-focused social companions.}
Deployments of ElliQ report reductions in loneliness and improved daily engagement when the device is proactive, uses short, forgiving dialogues, and hands off to phone or video channels when needed. Published design notes emphasise clear boundaries, avoiding overly human-like presentation, and multi-modal prompts, combining voice and visual cues, to support diverse users~\cite{broadbent2024elliq}. Studies show that anthropomorphic cues can both increase trust and trigger discomfort depending on context and user expectations~\cite{chen2025does}. Systems should therefore be proactive but bounded, announcing intentions, offering opt-outs, and handing over to human channels when system confidence is low.

\subsubsection{Telepresence platforms in aged care (e.g., temi robots).}
Early implementation and Quality Improvement (QI) reports from long-term care show resident interest in self-navigating telepresence for activities and check-ins, provided sessions are scheduled and content is staff-curated. The key technical needs are robust voice control, vital-sign connectors, and Electronic Health Record (EHR) integration~\cite{weigand2022concept}. A pragmatic starting point is a minimal clinical payload that includes device vitals and brief encounter notes, easing adoption while building toward deeper integration.

Section~\ref{Sec:Benchmark} revisits these interaction patterns and evaluates their real-world performance, using metrics such as dock success rate, false activations in speech, false alarms per day, and time to recover after faults, alongside usability and workload scores.

%-----------------------------------------------------------------------------------

%\clearpage
% --------------------------------------------------------------

\section{Technologies for perception and ambient systems}
\label{Sec:Technologies}

%------------------------
%************************
\subsection{Sensors and devices}
\label{subsec:sensors_technologies_chap5}

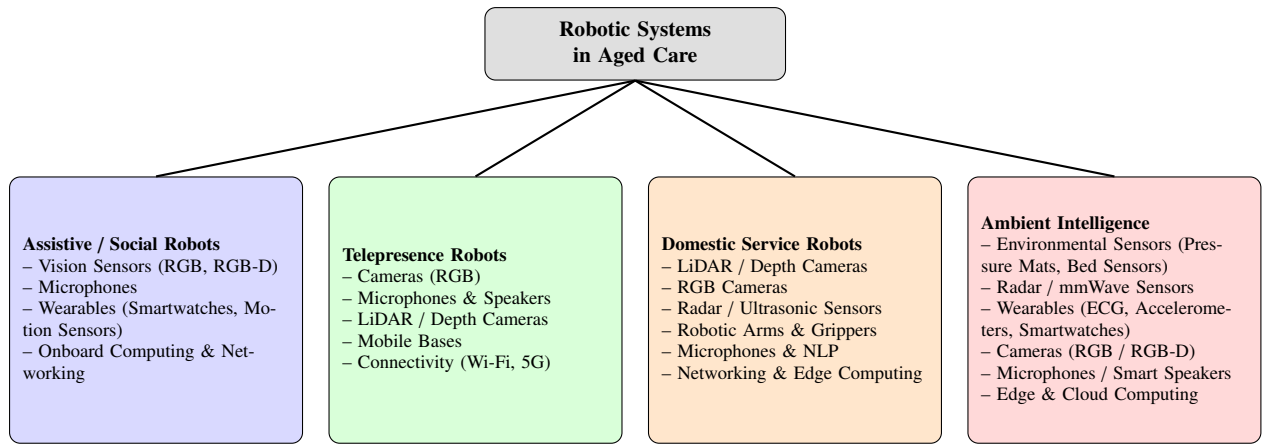
\begin{figure}[ht]
\centering
\resizebox{\linewidth}{!}{%
\begin{tikzpicture}[
  grow=down,
  level 1/.style={sibling distance=42mm, level distance=35mm},
  edge from parent/.style={draw, thick},
  edge from parent path={(\tikzparentnode.south) -- (\tikzchildnode.north)},
  parentbox/.style={
    rectangle, rounded corners, draw=black, align=center, font=\small\bfseries,
    text width=36mm, fill=gray!25, inner sep=5pt
  },
  box/.style={
    rectangle, rounded corners, draw=black, align=left, font=\scriptsize,
    text width=35mm, fill=gray!10, inner sep=5pt, minimum height=35mm % <-- fixed height for children
  }
]
%]

%\node[parentbox, fill=gray!25, font=\bfseries, text width=40mm, align=center] {Robotic Systems in Aged Care}
\node[parentbox] {Robotic Systems\\in Aged Care}
  child {node[box, fill=blue!15] {
    \textbf{Assistive / Social Robots}\\
    -- Vision Sensors (RGB, RGB-D)\\
    -- Microphones\\
    -- Wearables (Smartwatches, Motion Sensors)\\
    -- Onboard Computing \& Networking
  }}
  child {node[box, fill=green!15] {
    \textbf{Telepresence Robots}\\
    -- Cameras (RGB)\\
    -- Microphones \& Speakers\\
    -- LiDAR / Depth Cameras\\
    -- Mobile Bases\\
    -- Connectivity (Wi-Fi, 5G)
  }}
  child {node[box, fill=orange!20] {
    \textbf{Domestic Service Robots}\\
    -- LiDAR / Depth Cameras\\
    -- RGB Cameras\\
    -- Radar / Ultrasonic Sensors\\
    -- Robotic Arms \& Grippers\\
    -- Microphones \& NLP\\
    -- Networking \& Edge Computing
  }}
  child {node[box, fill=red!15] {
    \textbf{Ambient Intelligence}\\
    -- Environmental Sensors (Pressure Mats, Bed Sensors)\\
    -- Radar / mmWave Sensors\\
    -- Wearables (ECG, Accelerometers, Smartwatches)\\
    -- Cameras (RGB / RGB-D)\\
    -- Microphones / Smart Speakers\\
    -- Edge \& Cloud Computing
  }};
\end{tikzpicture}
}
\caption{Taxonomy of sensors and devices across robotic systems in aged care.}
\label{sensor_taxonomy}
\vspace{-6pt}
\end{figure}

Perception technologies are central to enabling robots and ambient systems to operate effectively in RAC and private homes. These environments impose unique constraints---cluttered layouts, variable lighting, mobility aids, and diverse sensory abilities---that shape sensing choices and integration strategies. Robust perception supports social engagement, health monitoring, safe navigation, and assistance with daily activities. The choice of sensors determines not only the functionality of these systems but also their acceptability, as issues of privacy, safety, cost, and usability are particularly sensitive in this context~\cite{costanzo2024new}.
Fig.~\ref{sensor_taxonomy} summarises the main sensor categories across four system types: 1) Assistive/social robots, 2) Telepresence robots, 3) Domestic service robots, and 4) Ambient intelligence. While this taxonomy is useful for mapping device types, the following discussion focuses on functional roles and design considerations across five themes: vision-based sensing, auditory and speech-based sensing, tactile and physical interaction sensing, physiological and environmental sensing, and enabling infrastructure.

%---------%--------------
\subsubsection{Vision-based systems}
Vision sensors form the perceptual foundation of robotic systems. They enable scene understanding \cite{fischinger2016hobbit}, object recognition\cite{fischinger2016hobbit}, navigation\cite{fischinger2016hobbit}, and social interaction \cite{mccoll2016survey}. They provide rich detail but raise strong privacy concerns, requiring safeguards such as restricted fields of view, on-device processing, and selective triggering. Vision sensors include:

\begin{itemize}
    \vspace{-6pt}
    \item \textbf{RGB cameras:}  
    RGB cameras are widely used in HRI systems. They enable functions such as face recognition \cite{fahn2022image}, activity recognition \cite{sawik2023robots}, gesture interaction \cite{sawik2023robots}, and fall detection \cite{benkaci2024vision}, which are critical in aged care. 
    Challenges include privacy risks from continuous recording, sensitivity to lighting changes, and high computational demands. 
    Despite these challenges, RGB cameras remain a foundational sensing modality across multiple robotic systems in aged care due to their versatility and rich visual information. Mitigation strategies include on-device analysis, privacy-preserving video summarisation, and sparse event triggers that limit continuous recording in private spaces.

    \vspace{-6pt}
    \item  \textbf{RGB-D cameras and depth cameras:} 
    RGB-D cameras combine colour and depth information to create a three-dimensional view of the environment, allowing robots to interpret spatial structures, recognise human postures, and navigate safely in cluttered or dynamic environments \cite{fanello2014learning,yang2020homecare}. In aged care, RGB-D sensors have been successfully applied to fall detection, posture estimation, and gesture-based human-robot interaction.
    Limitations include sensitivity to ambient lighting and reflective surfaces \cite{sarbolandi2015kinect} and increased computational demands. Although depth maps are less intrusive than RGB imagery, continuous 3D monitoring still raises privacy and acceptability concerns, and deployments often restrict depth capture to zones like beds or doorways.
    \vspace{-6pt}
    \item \textbf{LiDAR sensors:} 
    In aged care, LiDAR is primarily employed for safe robot navigation, environment mapping, and obstacle avoidance.
    Its ability to generate accurate spatial models independent of lighting conditions makes it well-suited for both day and night operation, overcoming some of the limitations of camera-based systems \cite{frovik2021utilising}.
    However, LiDAR sensors are typically more expensive and power-intensive, which limits their inclusion in compact or low-cost devices. They also lack semantic detail, necessitating fusion with vision or auditory data for full situational awareness.
\end{itemize}

%---------%--------------
\subsubsection{Auditory and speech-based sensing}
    Auditory sensors enable robots to perceive and interpret sound, speech, and tone, facilitating natural interactions. Microphones, speaker arrays, and smart audio systems are commonly embedded in social and telepresence robots to facilitate communication, emotional engagement, and safety monitoring \cite{christoforou2019overview,costanzo2024new}. While speech interfaces are intuitive, they remain vulnerable to noise and overlapping conversations. 

\begin{itemize}
    \vspace{-6pt}
    \item \textbf{Microphones:} 
    Microphones serve as the audio input in assistive robots \cite{fahn2022image}, capturing speech and environmental sounds for tasks such as voice navigation, speech recognition, and distress detection \cite{sawik2023robots}.
    Many systems use simple classifiers to detect salient events such as shouts, glass breaking, or repeated coughing, which can be performed at a lower computational cost and with reduced privacy exposure compared with continuous speech transcription.
    Challenges include background noise, multi-speaker interference, and continuous audio monitoring that can feel intrusive to residents and caregivers \cite{xu2025voice}. Designs that favour short buffered windows, on-device processing, and event summaries rather than raw audio streams can partially mitigate these concerns (see Section~\ref{Sec:Privacy}).
    
    \vspace{-6pt}
    \item \textbf{Speakers:} 
    Speakers enable robots to communicate verbally by delivering alerts, reminders, and navigation cues, as well as engaging in conversational interactions that support users’ mental and emotional well-being. Speech synthesis further facilitates naturalistic communication with residents, thereby improving both user acceptance and overall system usability~\cite{baltus2000towards}.
    However, speaker-based systems must be designed to avoid disruption or annoyance. Volume, tone, and clarity are critical considerations. Coordinating speech output with visual or contextual cues from other sensors ensures coherent communication. Ethical considerations include avoiding deceptive or overly persuasive speech and ensuring that users can control or mute robot communication as desired \cite{oliveira2017speaking}. In practice, acoustic design for prompts and alerts should follow the accessibility and escalation principles in Section~\ref{Sec:Interaction}.
\end{itemize}

%---------%--------------
\subsubsection{Tactile and physical interaction sensing}
Tactile and force sensors allow robots to physically interact with users and their surroundings safely. They are crucial for tactile communication and for applications such as robotic feeding and mobility assistance, where physical contact is integral to function \cite{topping2002overview}. Most current care home deployments involve limited or no physical manipulation, but safe touch sensing is a necessary condition for future assistance, for example, transfers, dressing, or feeding.

\begin{itemize}
    \vspace{-6pt}
    \item \textbf{Force and pressure sensors: } 
    Force and pressure sensors are critical for detecting contact forces during interaction with humans or objects. In aged care applications, they are integrated into robot joints, grippers, and support surfaces such as robotic arms, exoskeletons, or assistive beds to ensure gentle handling and to prevent injury \cite{popovic2019feeding}. Force and pressure sensors can be sensitive to calibration drift and mechanical wear, which may affect precision and reliability. Ensuring consistent performance requires careful sensor design, regular calibration, and the integration of redundant sensing modalities \cite{billard2019trends}. 
    \vspace{-6pt}
    \item \textbf{Artificial skins and touch sensors: } 
    Artificial skins are arrays of distributed sensors that mimic human skin, enabling robots to detect contact location, magnitude, and texture \cite{silvera2015artificial}. These sensors enhance safety by allowing robots to react immediately upon close proximity or touch, halting motion or adapting force before discomfort occurs. However, developing tactile materials and mechanisms that are durable, flexible, and sensitive remains an ongoing challenge. Many tactile sensors are prone to damage from wear, temperature changes, or humidity, limiting their long-term usability in domestic environments~\cite{silvera2015artificial,kim2022care}.
    \vspace{-6pt}
    \item \textbf{Proximity and contact sensors: }  
    Proximity sensors, such as infrared, capacitive, or ultrasonic sensors, allow robots to detect nearby obstacles and anticipate physical contact before it occurs \cite{topping2002overview}. In care robots, they are vital for collision avoidance, safe navigation around people, and pre-contact detection during assistive interactions. 
    While proximity sensors enhance safety, they often provide limited spatial resolution and may be influenced by environmental factors like lighting or surface reflectivity,  which necessitates combining proximity data with other complementary information \cite{sawik2023robots}.
\end{itemize}

%---------%--------------
\subsubsection{Physiological and environmental sensing} 
Wearables and ambient sensors monitor vital signs and environmental conditions, supporting context-aware assistance \cite{song2024emotion, lee2025impact}. While these sensors enhance safety, they introduce challenges around accuracy, calibration, and privacy. Data typically flows through local hubs that aggregate signals and trigger alerts without exposing raw streams.
    
\begin{itemize}
    \vspace{-6pt}
    \item \textbf{Physiological sensors:} 
    Physiological sensors measure biological signals such as heart rate, respiration, temperature or blood oxygen levels \cite{mukhopadhyay2007sensors}. These can be embedded in wearable devices or integrated directly into robotic systems. In aged care, physiological sensing enables real-time monitoring of vital signs to detect abnormalities such as cardiac irregularities or respiratory distress. 
    Challenges include reduced accuracy affected by motion, sweat, and inconsistent skin contact, while discomfort may reduce long-term use \cite{heikenfeld2018wearable}. Reliable data interpretation remains challenging due to substantial individual variability, and the use of such information raises significant privacy concerns given the sensitivity of the associated health data.
    
    \vspace{-6pt}
    \item \textbf{Environmental sensors:} 
    Devices tracking temperature, humidity, air quality, and light provide context for safe and adaptive behaviour  \cite{duthie2024design}. They complement physiological sensing by providing contextual awareness, but face challenges such as sensor calibration, drift, and false alarms caused by environmental fluctuations. Combining multiple streams of data requires sophisticated sensor fusion and decision-making frameworks to ensure reliable, actionable insights. Privacy concerns extend to continuous environmental monitoring, particularly when combined with vision or audio sensing that may inadvertently reveal personal habits or routines.
\end{itemize}

In practice, most sensing occurs through ambient or wearable devices rather than the robot alone, feeding into shared event and alert channels (Section~\ref{Sec:Interaction}), with the privacy implications discussed in Section~\ref{Sec:Privacy}.

%---------%--------------
\subsubsection{Technology enabling infrastructure}
Enabling infrastructure comprises the foundational systems that support sensing and robotic technologies, including IoT frameworks, wireless networks, edge and cloud computing, and data management architectures~\cite{costanzo2024new}. These components enable seamless coordination and responsive care environments. Architectural patterns should prioritise local inference, minimal data flows, and resilience under network loss, aligning with privacy-by-design principles discussed in Section ~\ref{Sec:Interaction}.

%------------------------
%************************
\subsection{Scene understanding and mapping}

Scene understanding and mapping are critical capabilities for robotic systems operating in RAC and home environments. These systems must navigate safely around residents, mobility aids, and personal belongings while performing tasks such as fetch-and-carry, monitoring, social interaction, and fall detection \cite{ni2023deep}. Unlike controlled laboratory settings, aged-care settings are highly dynamic and unpredictable environments, so robots require robust, context-aware scene understanding tailored to care scenarios rather than generic indoor navigation (see Section \ref{Sec:Background}).

Current approaches rely on multi-modal sensing using the devices described in Section \ref{subsec:sensors_technologies_chap5} to combine visual, depth, and non-visual information. RGB-D cameras are commonly used for capturing both appearance and geometry, supporting depth-based object detection and mapping. Complementary sensors such as inertial units, thermal cameras, radar, and ambient IoT devices add resilience when visual sensing is impaired \cite{duncan2024survey}. Beyond 2D perception, semantic simultaneous localisation and mapping (SLAM) integrates geometric mapping with object labels, allowing robots to reason about spatial relationships such as the proximity of a walking frame to a bed or the typical route between a resident’s chair and the bathroom~\cite{hing2020smart}.  

Despite recent progress, reliable scene understanding in care environments remains challenging because clutter, dynamic lighting, and frequent occlusions degrade perception~\cite{schuster2010perceiving,hung2022facilitators}. Layouts differ across homes, furniture and personal items move over time, and residents may use mobility aids or medical devices that complicate visual interpretation. Models trained on public datasets often fail to generalise without extensive adaptation, and long-term deployment introduces further complexity as robots must track gradual environmental and behavioural changes rather than assuming a static map.

Privacy introduces additional constraints \cite{rigaud2024implementation}. In this vein, privacy-preserving perception has become an increasingly active area of research within mapping and scene understanding \cite{taras2024inherently}. Many systems now avoid storing raw RGB data and instead rely on mapping data to privacy-preserving representations such as texture-free geometric maps, semantic maps without visual appearance, blurred or downsampled point clouds, and topological maps that encode only key locations (e.g., ``bed area'' or  ``bathroom entrance''), without detailed imagery \cite{taras2024inherently}. These maps allow robots to perform navigation and assistive tasks while reducing risks related to surveillance, consent, or data retention.

From a technical perspective, real-time processing on embedded hardware poses further limitations. Achieving accurate perception under power and latency constraints is challenging, especially when robots must process multi-modal data continuously in shared spaces. Efficient architectures, model compression, and on-device inference strategies are therefore essential for practical deployment \cite{rout2024development,zijiang2022mapping}.

%************************
\subsection{Human perception}
Robotic and intelligent systems provide a suite of capabilities that enable machines to interpret and respond to human presence, actions, and emotional states. Recent advances in computer vision and machine learning have significantly improved performance across detection, tracking, pose estimation, gesture recognition, fall detection, and affect recognition, making these technologies increasingly viable for assistive and aged care environments. 
    
\begin{itemize}
    \vspace{-6pt}
    \item \textbf{Human detection and tracking: } 
    Contemporary human detection and tracking systems leverage deep learning object detectors such as YOLO \cite{reis2023real} and transformer-based architectures \cite{meinhardt2022trackformer, gopal2024separable}, often integrated with multi-object tracking algorithms like ByteTrack \cite{zhang2022bytetrack} for robust performance in dynamic environments. These systems excel in multi-camera setups, enabling persistent tracking across viewpoints despite occlusions and illumination changes, and in care settings, they support tasks such as monitoring corridor use or checking whether a resident has returned to bed. LiDAR-assisted detection \cite{du2024lidar} and radar-based sensing are emerging as privacy-preserving alternatives, offering resilience under poor lighting and clutter. Challenges remain in achieving real-time performance on resource-constrained platforms and ensuring robustness under diverse environmental conditions.
    \vspace{-6pt}
    \item \textbf{Pose and gesture recognition: }
    Human pose estimation has evolved from 2D keypoint detection to 3D reconstruction using monocular RGB inputs \cite{yiannakides2019real}, multi-view systems, and hybrid sensor fusion \cite{bao2024hybrid}. Deep learning approaches incorporating convolutional neural networks \cite{luoorinet}, graph-based models  \cite{zou2021modulated}, and transformers \cite{fu2023deformer} dominate the field, with recent innovations such as diffusion models  \cite{cheng2024handdiff} and efficient tokenization strategies  \cite{li2024hourglass} reducing computational overhead while maintaining accuracy. These methods enable fine-grained gesture recognition critical for natural human-robot interaction. Despite progress, occlusion handling and generalisation to real-world scenarios remain open challenges, particularly in aged care settings where mobility aids and furniture introduce visual complexity and where residents may have a limited range of motion.
    \vspace{-6pt}
    \item \textbf{Fall detection: }
    Vision-based fall detection systems increasingly integrate pose estimation with anomaly detection frameworks to distinguish falls from normal activities. Transformer-based architectures \cite{al2024human, mccall2024computer} combined with lightweight deployment strategies (e.g., on edge devices) have demonstrated high accuracy while preserving privacy through local processing. Emerging approaches also explore multi-modal sensing, including radar and depth cameras \cite{cippitelli2017radar}, to overcome limitations of RGB imaging under occlusion or poor lighting. Nonetheless, dataset scarcity for real-world falls and the need for unobtrusive monitoring remain significant barriers to widespread adoption in care homes and private houses.
    \vspace{-6pt}
    \item \textbf{Affect recognition: }
    Affect recognition has progressed beyond facial expression analysis to incorporate body pose \cite{filntisis2019fusing}, contextual cues \cite{han2025salience}, and multi-modal fusion of visual, vocal, and physiological signals \cite{udahemuka2024multimodal}. Vision-language models \cite{bhattacharyya2025evaluating} and diffusion-based frameworks \cite{hassan2024self} now enable context-aware emotion inference, addressing limitations of earlier facial-only approaches. Micro-expression detection \cite{verburg2019micro} and transformer-based architectures \cite{zhang2022short, vazquez2022transformer} further enhance robustness in real-world conditions. However, challenges persist in data availability, cultural variability in emotional expression, and ethical considerations surrounding privacy and consent in continuous monitoring, especially when systems infer affect implicitly rather than through explicit check-ins.
\end{itemize}

%------------------------
%************************
\subsection{Learning paradigms in robotics for RAC and home care}

%---------%--------------
\subsubsection{Introduction to learning paradigms in HRI}
\begin{figure}
\centering
\includegraphics[width=0.85\linewidth]{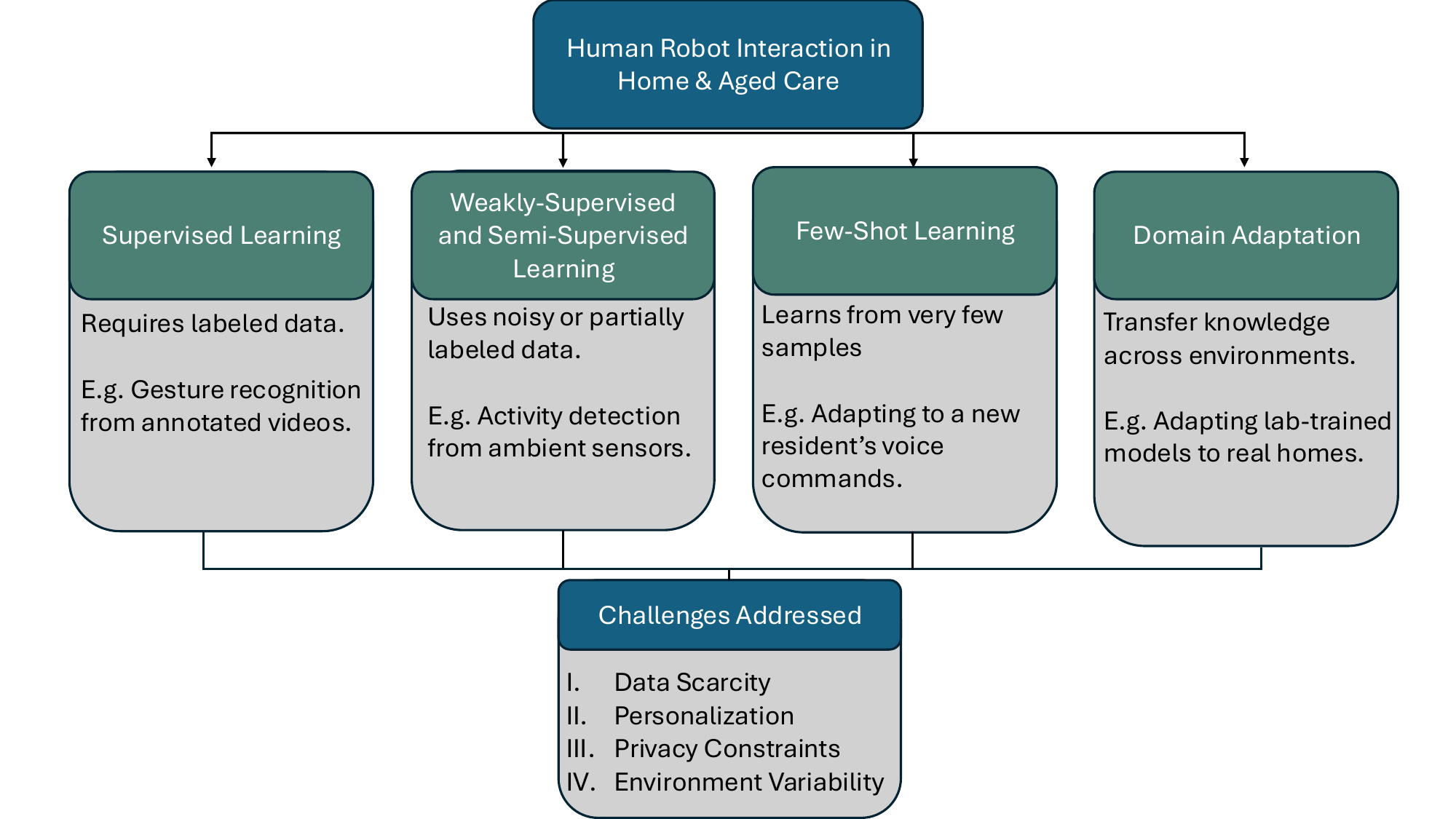}
\caption{Overview of the practical challenges of HRI in home and aged care, and how existing learning paradigms address these challenges.}
\label{fig:learning_1}
\vspace{-6pt}
\end{figure}

As assistive technologies enter home and RAC settings, robots must learn and adapt to dynamic, personalised environments. Learning paradigms, the frameworks through which robots acquire and refine knowledge, are central to this adaptability. These range from supervised learning, reliant on labelled data, to weakly-supervised, self-supervised, and few-shot approaches that reduce annotation needs. Domain adaptation enables models trained in controlled settings to function in diverse real homes. Fig.~\ref{fig:learning_1} illustrates these techniques and their challenges.

%---------%--------------
\subsubsection{Supervised learning approaches}
Supervised learning remains one of the most fundamental and widely used paradigms in robotics in care settings. It trains models on datasets where each input is paired with a corresponding label, enabling the system to learn a mapping from inputs to outputs. In typical pipelines, multimodal data streams (e.g., RGB video, audio, inertial data) are annotated with ground-truth labels such as activities, spoken commands, or gestures, and used to train deep architectures (e.g., CNNs, RNNs, Transformers) that minimise a task loss via back-propagation. Once validated on held-out data, these models are deployed and may be fine-tuned for specific homes or devices.

In home and RAC HRI, this pipeline underpins many core capabilities, for example, activity recognition for monitoring, speech recognition for voice commands, and gesture interpretation for safe interaction. Rather than detailing the generic pipeline, we focus on representative state-of-the-art extensions and applications that illustrate how these methods address care-specific challenges.

\begin{itemize}
    \vspace{-6pt}
    \item \textbf{Activity recognition:} Supervised learning enables robots to identify ADLs like eating or walking. Recent work uses multi-modal fusion of RGB, depth, and inertial data for robust classification. Robinson and Nejat \cite{robinson2023deep} introduced spatial mid-fusion for integrating geometric and semantic features, while RNNs mitigate observation noise \cite{mihoub2021deep}. Modern models like DeepConvLSTM and attention-based architectures improve temporal reasoning and interpretability in cluttered homes.
     \vspace{-6pt}
    \item \textbf{Speech recognition:} Transformer-based ASR systems such as Conformer \cite{gulati2020conformer} and Whisper \cite{radford2023robust} achieve state-of-the-art by combining convolution and self-attention. Parameter-efficient prompt tuning \cite{ma2024extending} enables personalisation with minimal overhead. Domain-specific fine-tuning and end-to-end pipelines improve robustness in home-care contexts \cite{pallavoor2024gesture}.
     \vspace{-6pt}
    \item \textbf{Gesture recognition:} Shifted from rule-based to deep learning using RGB-D cameras and graph neural networks \cite{swietlicka2025survey}. DiG-Net \cite{beeri2025dig} achieves 97.3\% accuracy with depth-conditioned alignment for long-range gestures, while GestLLM \cite{kobzarev2025gestllm} combines MediaPipe features with LLMs to interpret culturally nuanced gestures without retraining.
\end{itemize}

Despite its widespread success, supervised learning faces major hurdles in home and RAC robotics. Data annotation is a key barrier: creating labelled datasets is costly and time-intensive. Annotating RGB-D video, LiDAR, and audio streams demands expertise and specialised tools, especially for segmenting continuous video and verifying object interactions. Domain mismatch compounds this challenge, with models trained in labs often failing in the real world.
    
Generalisation remains limited. Supervised models tend to overfit to training data and struggle with rare activities or atypical gestures, or speech patterns common in older adults. Finally, scalability and privacy introduce ethical constraints. Collecting large-scale labelled data in private homes conflicts with autonomy and data protection principles, making continuous monitoring impractical and reinforcing the need for privacy-preserving alternatives.

%---------%--------------
\subsubsection{Weakly-supervised and self-supervised learning approaches}

Traditional supervised learning relies on large, fully annotated datasets, which are expensive and time-consuming to create. \textit{Weakly-supervised learning (WSL)} relaxes this requirement by using incomplete, inexact, or noisy labels, for example, coarse timestamps or video-level tags instead of frame-level annotations. \textit{Self-supervised learning (SSL)} goes a step further by eliminating manual labels entirely. It leverages the intrinsic structure of the data to create pseudo-labels or predictive tasks, such as predicting missing frames in a video or reconstructing masked audio segments. Both paradigms aim to exploit unlabelled or weakly labelled data that can be collected passively, reducing annotation costs.

WSL has gained traction in HRI for tasks where fine-grained annotation is impractical. For example, in action segmentation, \citet{yang2023weakly} introduced Action-Union Learning, which uses timestamp-level annotations instead of frame-wise labels, achieving near-supervised performance with less than 1\% of full labels. In object detection, \citet{maiettini2021weakly} proposed a teacher-learner pipeline for humanoid robots, combining weak supervision with online learning to adapt detection models in real time, reducing labelling effort while leveraging robot usage for active data acquisition. Similarly, domain adaptation with weak supervision has been explored by \citet{wilson2020multi}, who developed CoDATS, a multi-source adaptation framework for time-series sensor data that incorporates weak supervision signals such as label distributions, improving activity recognition across heterogeneous home environments. Finally, for activity recognition, semantic-guided attention mechanisms have been used to eliminate the need for bounding box annotations and reduce computational overhead \cite{liu2023semantic}.

SSL enables robots to learn rich representations from raw sensory streams without manual labels, making it highly relevant for HRI in low-resource and privacy-sensitive settings. In speech and multi-modal interaction, SSL has transformed automatic speech recognition (ASR) through models such as wav2vec 2.0 and HuBERT, which learn robust audio embeddings by predicting masked segments, significantly reducing dependence on labelled speech data. These embeddings are now integrated into multi-modal HRI systems for voice-enabled assistive robots \cite{mohamed2022self}. Similarly, \citet{pan2022leveraging} leveraged unimodal SSL for audio-visual speech recognition, pretraining audio and visual encoders separately and then fusing them for lip-reading tasks, improving robustness in noisy environments where visual cues complement degraded audio. Beyond speech, SSL has advanced robotic manipulation and self-modelling: \citet{florence2019self} introduced a video prediction-based approach that enables robots to anticipate object trajectories and arm movements, supporting zero-shot transfer to new tasks in home environments. Likewise, \citet{mezghani2023learning} demonstrated goal-conditioned SSL for manipulation, allowing robots to infer task goals from visual cues and adapt actions accordingly, reducing reliance on human-provided demonstrations.

Despite their promise, WSL and SSL have notable limitations. WSL often relies on heuristic labels or coarse annotations, which can introduce noise and ambiguity; models may overfit to incorrect patterns, reducing reliability in safety-critical assistive tasks. It also does not eliminate the need for domain adaptation; models trained on partially labelled datasets may still struggle with cultural and environmental variability in homes. 

SSL, while reducing annotation costs, introduces different challenges. Designing effective pretext tasks that transfer well to downstream HRI applications remains difficult, since predicting masked video frames or audio segments may not directly improve gesture recognition or activity classification without extensive fine-tuning. SSL methods also demand significant computational resources for large-scale pretraining, which may be impractical for resource-constrained robots. Both paradigms share issues of limited interpretability, raising concerns about transparency and trust in care contexts. Finally, collecting raw sensory data for pretraining raises privacy-related ethical concerns.

\subsubsection{Few-shot and meta-learning approaches}
In RAC and home care, robots must adapt quickly to new users, layouts, and routines with only a handful of examples. Unlike traditional systems needing large datasets, few-shot and meta-learning enable rapid adaptation with minimal input. Few-shot learning generalises from limited examples, ideal for scenarios where collecting extensive data is impractical or intrusive. A robot might learn routines, dietary needs, or communication styles from just a few interactions, supporting personalised assistance and trust. Compared to supervised learning, which demands labelled data, few-shot learning is inherently suited to data-scarce environments.

Few-shot learning is most useful during onboarding or change, when residents move rooms or adopt new routines, allowing robots to update behaviours without full retraining. Meta-learning underpins this capability by teaching robots how to adapt across tasks rather than memorising routines. Frameworks like MAML and Prototypical Networks train on diverse scenarios, enabling quick fine-tuning in real homes. In this chapter, meta-learning is considered not as an abstract concept but as a practical tool for safe, low-effort personalisation across varied homes, languages, and care routines.

Few-shot learning has advanced significantly for aged care applications. The VITAL framework \cite{kasaei2024vital} combines few-shot imitation learning with human-in-the-loop corrections, enabling robots to learn tasks such as setting trays or stacking objects from only a few demonstrations, augmented by simulation and feedback. Similarly, Wi-Fi-based sensing \cite{wang2024review} applies few-shot learning to detect activities like falls without intrusive sensors, supporting privacy-preserving monitoring. For personalisation, preference-based action representation \cite{wang2025personalization} allows robots to fine-tune behaviours—such as feeding angles or tone—from minimal user feedback. Meta-learning further drives cross-task adaptability: approaches like variational meta-RL \cite{ballou2023variational} help robots adjust to different reward functions, while trust modelling \cite{gao2019fast} fosters rapport in long-term care. Meta-imitation learning enables skill generalisation from limited demonstrations, supported by benchmarks such as Meta-World \cite{yu2020meta}. Finally, meta-learning integrates with adaptive control for fine motor tasks, as in peg-in-hole assembly \cite{johannsmeier2019framework}, combining skill formalism with meta-parameter learning for precision tasks like assistive device maintenance.

Despite their promise, few-shot and meta-learning approaches face several challenges. First, task diversity and complexity can limit generalisation: while these methods excel at adapting to simple routines or object manipulation, highly variable activities (e.g., culturally nuanced interactions) remain difficult to learn from only a few examples.  Robustness and interpretability are also concerns; systems can be brittle and opaque, limiting caregiver oversight. Moreover, meta-learning adds complexity: training across diverse tasks demands large datasets or simulations, which are costly. Models may overfit to seen tasks and fail when behaviours differ, as assumptions about shared structure often break in personalised care. Both paradigms rely on limited or implicit feedback, yet older adults may provide unclear signals, and caregivers lack time for fine-tuning. This gap between theory and practice raises ethical risks: incorrect adaptation could harm or undermine dignity.

In summary, while promising for personalisation, few-shot and meta-learning face hurdles in data quality, robustness, interpretability, computational cost, and ethics that must be addressed for real-world deployment.

%---------%--------------
\subsubsection{Domain adaptation for home environments}

As robots move from lab settings to home and RAC settings, they face a number of challenges. Domain adaptation (DA) addresses these gaps so models trained in one domain (e.g., simulation or benchmark datasets) remain reliable in real-world settings. Unlike supervised learning, which assumes independent and identically distributed (i.i.d.) data, DA explicitly handles covariate and conditional shifts, enabling robust perception, manipulation, and interaction under variability \cite{hassanpour2023survey}. In home-care, DA reduces costly data collection and relabelling, improving safety and consistency.

Core DA methods learn domain-invariant yet task-relevant representations. Techniques include discrepancy minimisation (e.g., MMD), adversarial alignment (e.g., DANN), and self-training with pseudo-labels, as well as active-learning loops that selectively query caregivers or clinicians for labels on uncertain or safety-critical events. 
For example, adversarial DA confuses a domain discriminator to strip domain-specific cues (e.g., furniture textures) while preserving semantics like object categories or grasp points. These approaches support reliable activity recognition and control across diverse homes. Additionally, compared to supervised learning, which degrades under target shift, DA is purpose-built for cross-domain generalisation. SSL and WSL reduce annotation needs but do not guarantee invariance. DA directly targets distribution mismatch, making it well suited for highly variable HRI environments and personalised care contexts \cite{hassanpour2023survey}.

Recent DA research focuses on embodied AI and home robotics, where sim-to-real transfer and non-stationary targets are common. Simulators such as Habitat and iGibson, along with photorealistic datasets like Matterport3D and Replica, enable large-scale source training, while DA mitigates real-world gaps through feature alignment, style transfer, and test-time adaptation \cite{savva2019habitat, li2021igibson}. For manipulation tasks, sim-to-real pipelines incorporate dynamics randomisation and visual perturbations to improve pick-and-place and assistive behaviours despite household variability \cite{peng2018sim}. Adversarial DA methods (e.g., DANN, ADDA) combined with multi-target adaptation help avoid catastrophic interference \cite{hassanpour2023survey}, while test-time adaptation \cite{wang2022continual} and batch-normalisation alignment \cite{chang2019domain} enable on-the-fly adjustments to lighting or sensor drift. Discrepancy-based DA (e.g., MMD) and pseudo-label self-training further enhance segmentation of cluttered scenes for safe navigation \cite{zou2018unsupervised}. 

Beyond perception, reinforcement learning is increasingly combined with DA for policy-level adaptation and personalisation over time. In home and aged-care deployments, RL policies can adjust reminder timing, navigation strategies, or escalation thresholds based on longitudinal feedback from residents and staff, effectively treating changes in routines, mobility, or furniture layouts as a form of temporal domain shift. Meta-reinforcement learning approaches for social robotics and robotic skill adaptation show how policies can be quickly re-tuned when reward functions or dynamics change across environments. Examples of such an approach include variational meta-RL for social robots~\cite{finn2017model,zintgraf2019varibad} and variational information-bottleneck regularised deep RL for robotic skills adaptation~\cite{xiang2023variational}.

Vision–language models for household tasks (e.g., ``find medicine box'') also benefit from DA across viewpoints and linguistic variations, improving instruction following without exhaustive retraining. Combined with few-shot personalisation and active learning during deployment, DA supports rapid adaptation to user routines while maintaining robustness. 

Despite its advantages, domain adaptation faces several challenges.  Negative transfer can occur when source and target distributions differ substantially (e.g., minimal overlap between training homes and a target residence with unique layouts or cultural artifacts). Without reliable measures of transferability, alignment may degrade task performance or induce spurious invariances \cite{ng2024improving}. 
Similarly, adversarial alignment can also overemphasise domain indistinguishability at the expense of fine-grained semantics, harming safety-critical tasks such as medication recognition or fall-risk area detection in cluttered rooms \cite{wang2020classes, zhu2024adversarially}. 

Label scarcity and annotation noise remain problematic: pseudo-labels can propagate errors when target conditions are highly variable (moving furniture, seasonal lighting changes, sensor ageing). 
Robust self-training, uncertainty estimation, and curriculum-style adaptation are needed to mitigate confirmation bias \cite{huang2024dtbs, ko2025active}. 

For long-term deployment, RL-based DA introduces additional difficulties: reward signals may be sparse or delayed, user preferences and mobility can drift over months, and safe exploration is heavily constrained in care settings. Practical personalisation therefore relies on conservative update rules, explicit human-in-the-loop feedback, and continual-learning schemes that view temporal drift as a sequence of related domains rather than a single stationary environment~\cite{garcia2015comprehensive,liu2022robustness}.

Finally, in RAC and private homes, these methods must also operate under strict privacy and consent constraints, further limiting the volume and granularity of target data available for adaptation and reinforcing the need for conservative, well-audited DA strategies.

%-----------------------------------------------------------------------------------

%\clearpage
% --------------------------------------------------------------
\section{Privacy, trust and acceptability}
\label{Sec:Privacy}

Privacy, trust, and acceptability are foundational requirements for deploying robots and ambient intelligence in private and RAC homes.
These systems operate in intimate spaces, observing sensitive routines and often supporting people with reduced capacity to monitor or control technology, so they must protect dignity, choice, and control, not just deliver useful support.
As a result, privacy expectations are not abstract ethical ideals but concrete design constraints that shape sensing, data flows, and interaction patterns. Trust shapes whether residents, families, and staff feel confident in the system, combining cognitive factors (e.g., reliability) and affective factors (e.g., perceived intentions and emotional connection)~\cite{rossi2021trust}. Acceptability refers to whether users perceive the system as suitable and useful in their context, balancing functional utility and emotional comfort~\cite{felding2023scoping,zhao2025social}. 
These concepts are tightly linked: ambiguous privacy erodes trust, and low trust can lead to rejection even when systems are technically capable.

Building on the constraints outlined in Section~\ref{Sec:Background} and the sensing and learning choices in Section~\ref{Sec:Technologies}, this section examines how these capabilities translate into practical safeguards, drawing on older adults’ perspectives on privacy and governance~\cite{slane2025bringing} and ethics-of-care frameworks that emphasise relational dynamics and vulnerability in aged care settings~\cite{yuan2023ethical}.
Together, these perspectives define a practical goal for this chapter: to advocate for systems that are useful, respectful, and explainable in the environments where people live.

Within this section, we outline the core privacy risks and corresponding design responses, the mechanisms that build and calibrate trust, and the factors that sustain acceptance over time. 
Users need clarity about what the system senses, when it is active, and how decisions are made; 
families and clinicians must be able to rely on alerts without the burden of frequent false alarms; 
and facility operators must ensure compliance with regional legal and policy frameworks. 
These themes connect directly to the interaction patterns in Section~\ref{Sec:Interaction} and underpin the evaluation lenses explored in Section~\ref{Sec:Benchmark}. 
The discussion is informed by recent studies on safe interaction methods~\cite{zhao2023research}, utility and acceptability in home care~\cite{zhao2025social}, and systematic reviews of assistive robot acceptance~\cite{jones2025acceptance}. 

%-----------------
\subsection{Privacy}

\subsubsection{Conceptual overview}
Privacy in socially-assistive robots and smart home systems has three dimensions. 
First, \textit{informational privacy}, or control over personal data such as faces, voices, and routines. 
Second, \textit{contextual privacy} which requires respecting social norms in sensitive spaces like bedrooms, bathrooms, or shared lounges. 
Third, \textit{decisional privacy}, the right to refuse, pause, or mute a device. For older adults, these concerns surface in clear questions: What does the robot record? Who will see it? How can I stop it~\cite{slane2025bringing}?

\subsubsection{Challenges}
Common risks in home care settings include always-on microphones, video recording, and silent background uploads. Consent is rarely a one-time event; cognitive changes, rotating staff, and visiting family complicate ongoing agreement. Some people feel safer with monitoring and accept intrusion, while others reject sensing even when it could help. This is the privacy paradox. Designs should respect both choices by making settings visible and reversible \cite{yuan2023ethical}. In practice, this means avoiding hidden sensing, setting conservative defaults, and allowing simple ways to pause or narrow sensing scope, all aligned with the interaction hooks described in Section~\ref{Sec:Interaction}.

\subsubsection{User perspectives}
Studies with older adults report a consistent trade-off: people value safety through fall detection or check-in calls, but resist constant recording. 
Comfort improves when systems provide visible states, such as a light that shows when audio is live, short default retention periods, and clear explanations. 
People also value ways to review what was captured and to delete it \cite{zhao2025social,slane2025bringing}. These expectations are similar in both RAC and private homes, though shared settings raise additional governance questions: who decides the rules, and how will disagreements be resolved?

\subsubsection{Policy and legal frameworks}
Teams should align with regional rules without reducing design to a compliance exercise. In the European Union, the AI Act sets risk-based obligations for systems that influence health and safety, including requirements for risk management, data governance, transparency, and human oversight \cite{euaiact2024}. ISO and IEC provide complementary guidance on managing risk in artificial intelligence systems~\cite{iso23894}. 
In Australia, the Office of the Australian Information Commissioner explains duties for collection, consent, access, and security for personal information, including audio and video~\cite{oaicAPPs2022}. Accessibility guidance such as WCAG 2.2 ensures controls remain visible and usable for people with low vision or hearing loss~\cite{W3CWCAG22}. These frameworks establish baselines for consent, logging, and access that technical and clinical teams can share.

\subsubsection{Privacy-by-design}
Privacy by design translates principles into concrete safeguards. Key practices include minimising data collection, preferring on-device inference, storing summaries instead of raw audio or video, applying face and voice blurring for data leaving the device, limiting retention with visible settings, and providing role-based access so carers see only what they need.  
A simple pipeline illustrates this: sense only what is needed, process data locally where possible, convert signals into filtered events (e.g., suspected falls or missed medication prompts), share events only with consented parties, and delete underlying data on a clear schedule. 
Privacy-preserving learning methods, such as federated learning and related approaches, can improve models without pooling raw personal data ~\cite{casado2024towards}. Transparency should be embedded, with live indicators when sensing is active and an easy one-touch pause~\cite{marchang2023security}. These ideas align with interaction hooks in Section~\ref{Sec:Interaction}, and uphold the principle that people should always be able to see and control what the system is doing.

Threat models for home robots and ambient sensors highlight risks such as unauthorised access, inference attacks, and silent data leakage~\cite{hung2022technological,zieni2025interdisciplinary}. These risks should be mitigated through conservative defaults, secure update mechanisms, and clear incident reporting paths that facility operators can follow when something goes wrong.

%-----------------
\subsection{Trust}

\subsubsection{Definition and dimensions}
Trust in human-robot interaction has two key dimensions: \textit{performance-based trust} reflects confidence that the system works reliably, while \textit{relational trust} reflects confidence that the system respects the person and will not deceive or harm them. In both private and RAC homes, these dimensions are inseparable: a robot that is accurate but opaque will not be welcome, and a robot that is friendly but unreliable will be ignored.

\subsubsection{Design considerations}
Ethical frameworks such as care-centred and value-sensitive design offer practical guidance for trustworthy HRI, emphasising transparency, autonomy, and emotional safety. Systems should clearly communicate what they sense and how decisions are made, while offering control and reversibility. When confidence is low, such as during fall detection or ambiguous commands, the system should defer to a human rather than act autonomously, as also reflected in recent reviews~\cite{yuan2023ethical,rossi2021trust}. Designers should avoid overstating capabilities and instead calibrate expectations through simple feedback, for example, by showing what the robot saw and why it responded. These principles align with fallback and recovery strategies discussed in Section~\ref{Sec:Interaction} and support relational trust, where users feel respected and safe.

\subsubsection{Interaction factors}
Three design choices shape trust in daily use. First, ensure operation transparent: indicate when microphones or cameras are active, and announce state changes with both light and speech so people with low vision or hearing can follow. Second, prioritise reliability and graceful recovery: apply the recovery ladder from Section~\ref{Sec:Interaction} with a Plan B that falls back to a large-screen prompt and a Plan R that retries before offering handover to a human contact. Third, aim for emotional fit: avoid human-like imitation that feels deceptive or stigmatising; instead, use a consistent voice, gentle motion, and simple expressive lighting that users can understand \cite{rossi2021trust,zhao2025social}. Trust is built gradually through honest behaviour, clear explanation of faults, and visible learning from incidents.

\subsubsection{Stakeholder collaboration}
Trust must be established with all stakeholders. Residents and families need clear consent and control. Care workers need actionable alerts and simple tools to adjust routines. Facility operators need logs and reports. Designers and vendors need ways to learn from field data without exposing personal information. Effective projects use short co-design cycles with staff and residents, then expand once routines are stable \cite{zhao2023research}. Longitudinal studies show that trust evolves over time and must be maintained through consistent behaviour and transparent updates~\cite{van2020designing,campagna2025systematic}, motivating the longitudinal evaluation measures discussed in Section~\ref{Sec:Benchmark}.

%-----------------
\subsection{Acceptability}

\subsubsection{Frameworks}
Acceptability in RAC and private homes is commonly assessed with two families of models. The Theoretical Framework of Acceptability (TFA) comprises seven constructs---affective attitude, burden, perceived effectiveness, ethicality, intervention coherence, opportunity cost, and self-efficacy---and has been recently applied to social robots supporting community-dwelling older adults~\cite{zhao2025social}. The Almere Model, adapted from the Technology Acceptance Model (TAM) and the Unified Theory of Acceptance and Use of Technology (UTAUT), extends these ideas with social-interaction factors such as trust, social presence, and perceived sociability, as demonstrated in nursing-home deployments~\cite{felding2023scoping}. 

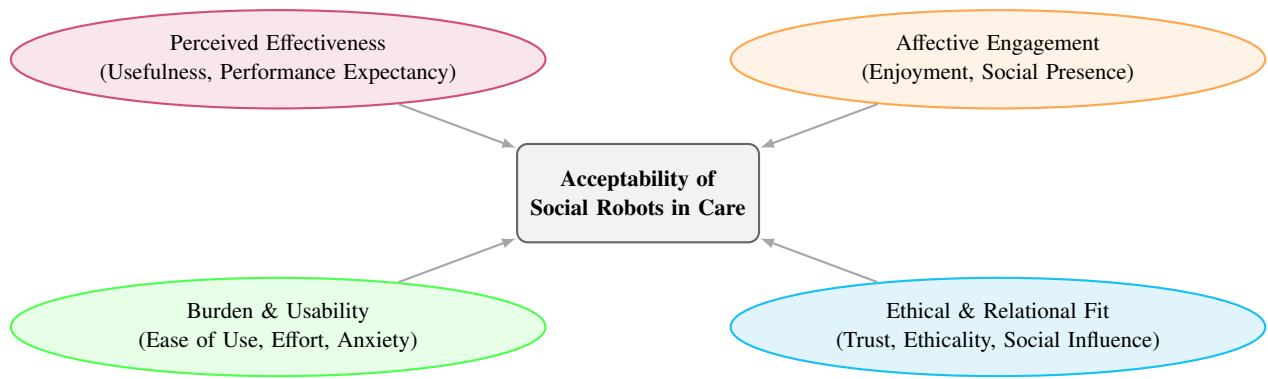
\begin{figure}[!t]
\centering
\begin{tikzpicture}[
    node distance=0.9cm, >=latex, thick,
    dim/.style={ellipse, draw=blue!60, fill=blue!10, text width=4.8cm, align=center, minimum height=1.3cm},
    framework/.style={rectangle, rounded corners, draw=black!60, fill=gray!10, text width=3cm, align=center, minimum height=1.3cm}
]

% Central node
\node[framework, font=\bfseries] (accept) {Acceptability of Social Robots in Care};

% Four domains
\node[dim, above left=of accept, fill=purple!10, draw=purple!70] (effectiveness) 
    {\hyphenpenalty=10000 Perceived Effectiveness\\(Usefulness, Performance Expectancy)};
\node[dim, above right=of accept, fill=orange!10, draw=orange!70] (affective) 
    {Affective Engagement\\(Enjoyment, Social Presence)};
\node[dim, below left=of accept, fill=green!10, draw=green!70] (burden) 
    {Burden \& Usability\\(Ease of Use, Effort, Anxiety)};
\node[dim, below right=of accept, fill=cyan!10, draw=cyan!70] (ethical) 
    {Ethical \& Relational Fit\\(Trust, Ethicality, Social Influence)};

% Connections
\draw[->, thick, gray!70] (effectiveness) -- (accept);
\draw[->, thick, gray!70] (affective) -- (accept);
\draw[->, thick, gray!70] (burden) -- (accept);
\draw[->, thick, gray!70] (ethical) -- (accept);

\end{tikzpicture}
\caption{
Integrative acceptability framework for social robots in RAC and private homes. The four domains perceived effectiveness, burden and usability, 
ethical and relational fit, and affective engagement represent recurring constructs shared across the TFA, Almere, and TAM/UTAUT model families~\cite{zhao2025social,felber2024informing,felding2023scoping}.
}
\label{fig:acceptance_framework}
\end{figure}

Qualitative extensions further highlight reliability, anxiety, and relational aspects in caregiver–user contexts~\cite{felber2024informing}. 
Across these traditions, acceptance constructs converge around four recurrent domains: 
(1) perceived effectiveness, 
(2)burden and usability, 
(3) ethical and relational fit, and 
(4) affective engagement, which together provide a unifying lens for evaluating social robots in care. 
These relationships are summarised conceptually in Fig.~\ref{fig:acceptance_framework}, which illustrates the four recurrent domains and their theoretical roots in the TFA, Almere, and TAM/UTAUT model families.

% ==============================
%  Diagram: integrative model
%==============================

% -----------------
\subsubsection{User studies}
Community trials show that acceptance improves when benefits are immediate and visible. For example, features such as easier family contact, routine reminders, and fall alerts that work without constant recording. Acceptance improves when the system is proactive but bounded, and when users can opt out at any time \cite{zhao2025social}. Reviews also stress clear design rules for contact and navigation in tight spaces \cite{zhao2023research}. Negative experiences often relate less to raw accuracy and more to frustration, interruption, or feelings of being monitored, which underscores the importance of the privacy and trust controls above.

\subsubsection{Influencing factors}
Acceptance is shaped by culture, language, and prior experience. Stigma is a real risk if a device looks like a medical alarm. Simple and friendly shapes fare better than humanoid imitations. Multilingual prompts and easy name customisation help in diverse homes. 
Training for staff and family is critical: acceptance grows when carers can adjust schedules and content without technical help \cite{jones2025acceptance}. Designers should therefore plan for configuration tools that fit care workflows, rather than assuming a technician will always be available.

\subsubsection{Long-term engagement}
Many projects experience a novelty effect in the first weeks, followed by a decline. Sustained use requires measuring daily participation over months and refreshing content regularly. Some people report discomfort when robots try to mimic humans too closely. This uncanny valley effect can be reduced with honest design that avoids human imitation and by setting clear boundaries on system capabilities \cite{rossi2021trust}. Ultimately, robots should be treated as part of a care program led by people, not as replacements.

\subsubsection{Co-design methods}
Participatory design approaches involving older adults and caregivers are essential for tailoring systems to real needs and improving long-term acceptability. Co-design helps uncover subtle preferences, cultural norms, and usability barriers that standard testing may miss. Studies show that early user involvement leads to more trusted and usable systems~\cite{sumner2021co,backaberg2024codesign}. These methods also support emotional engagement and ownership, which are key to sustained use in domestic and care settings. Co-design activities can also uncover privacy and trust expectations early, avoiding costly redesign when deployments reach the field.

%--------------
\subsection{Intersections and trade-offs}
Privacy, trust, and acceptability are independent but interconnected. Strong privacy defaults can reduce data for learning, which may lower performance, while aggressive automation may boost short-term efficiency but erode trust if people feel excluded from decisions. The balance depends on context: in a private home, a person may choose stronger privacy and accept fewer features; in an RAC, shared rules may be needed for fairness and consistency across residents.

These balances also differ across stakeholder groups. Older adults often prioritise agency, independence, and dignity even when this entails some residual risk; whereas clinicians and care staff emphasise safety, reliability, and efficiency within constrained workflows. Families may value reassurance and frequent updates, sometimes favouring more intensive monitoring than the resident themselves would prefer. Field studies in care homes repeatedly report such tensions between resident-centred and staff-centred goals, and show how they shape the configuration and day-to-day use of robots and ambient systems~\cite{de2019would,sharkey2012granny,yeo2023unpacking,carros2022care}.

A practical path emerges from earlier sections: make sensing and state visible, keep data local, offer simple explanations and controls, and evaluate beyond accuracy by considering burden, recovery, and long-term acceptance~\cite{slane2025bringing,yuan2023ethical,rossi2021trust}.
In practice, handling conflicting functional and ethical requirements means making these trade-offs explicit and negotiated rather than implicit. Co-design and governance processes should surface where resident agency, staff workload, clinical risk, and family reassurance pull in different directions; and the use of a simple task, risk, and burden lens to justify which compromises are adopted in a given deployment should be considered~\cite{sumner2021co}. 
Privacy trade-offs become concrete when teams decide what to log, how long to keep logged data, and which errors they are willing to tolerate, which is why Section~\ref{Sec:Benchmark} treats evaluation as part of governance rather than a separate technical step.

\subsection{Risk management and governance}
Risk management for RAC and private homes should connect design choices to governance duties and daily practice. It begins by defining the intended use and context, and then identifying the main classes of risk, such as physical safety, data protection, and psychosocial or dignity impacts. Controls should be simple and auditable: local inference where possible, short retention with clear settings, visible states for listening and recording, and conservative motion near people. Plans must include recovery from common faults (e.g., low battery, network loss, or obstructed sensors) and specify how human oversight intervenes when confidence is low. During deployment, teams monitor and report a small set of field metrics that matter to residents and staff, including near misses, false alarms per day, docking success, and time to recover after faults. These should be reviewed with caregivers at regular intervals.  A brief, regularly updated risk register helps track ownership and changes. This approach aligns with the European Union Artificial Intelligence Act and with ISO or IEC guidance on risk management for artificial intelligence, and helps prepare evidence for privacy and accessibility expectations in local regulations and standards \cite{euaiact2024,iso23894,oaicAPPs2022,W3CWCAG22}.

%-----------------------------------------------------------------------------------

%\clearpage
% --------------------------------------------------------------
\section{Benchmarking and evaluation}
\label{Sec:Benchmark}
RAC homes and private homes are dynamic, cluttered, and resource-constrained environments. Systems that are designed and evaluated only in labs often fail to capture what matters in daily use, including alarm fatigue, recovery after faults, caregiver workload, and dignity or privacy trade-offs. Benchmarking must therefore move beyond accuracy to include operational, ethical, and usability dimensions that reflect real-world conditions.

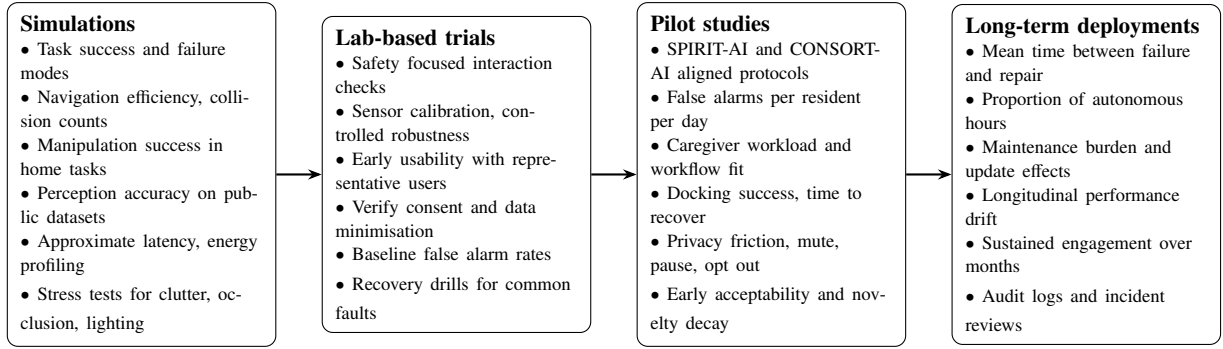
\begin{figure}[!t]
\centering
\begin{tikzpicture}[
    box/.style={
        draw,
        rounded corners,
        align=left,
        text width=3.2cm,
        inner sep=5pt
    },
    arrow/.style={-{Stealth[length=1.8mm]}, thick}
]

\node[box] (sim) {
\textbf{Simulations}\par
\footnotesize
\textbullet\ Task success and failure modes\par
\textbullet\ Navigation efficiency, collision counts\par
\textbullet\ Manipulation success in home tasks\par
\textbullet\ Perception accuracy on public datasets\par
\textbullet\ Approximate latency, energy profiling\par
\textbullet\ Stress tests for clutter, occlusion, lighting
};

\node[box, right=6mm of sim] (lab) {
\textbf{Lab-based trials}\par
\footnotesize
\textbullet\ Safety focused interaction checks\par
\textbullet\ Sensor calibration, controlled robustness\par
\textbullet\ Early usability with representative users\par
\textbullet\ Verify consent and data minimisation\par
\textbullet\ Baseline false alarm rates\par
\textbullet\ Recovery drills for common faults
};

\node[box, right=6mm of lab] (pilot) {
\textbf{Pilot studies}\par
\footnotesize
\textbullet\ SPIRIT-AI and CONSORT-AI aligned protocols\par
\textbullet\ False alarms per resident per day\par
\textbullet\ Caregiver workload and workflow fit\par
\textbullet\ Docking success, time to recover\par
\textbullet\ Privacy friction, mute, pause, opt out\par
\textbullet\ Early acceptability and novelty decay
};

\node[box, right=6mm of pilot] (long) {
\textbf{Long-term deployments}\par
\footnotesize
\textbullet\ Mean time between failure and repair\par
\textbullet\ Proportion of autonomous hours\par
\textbullet\ Maintenance burden and update effects\par
\textbullet\ Longitudinal performance drift\par
\textbullet\ Sustained engagement over months\par
\textbullet\ Audit logs and incident reviews
};

\draw[arrow] (sim) -- (lab);
\draw[arrow] (lab) -- (pilot);
\draw[arrow] (pilot) -- (long);

\end{tikzpicture}
\caption{A staged evaluation timeline for assistive and social robots and ambient AI in private and care homes. The emphasis shifts from dataset and simulator metrics to safety, workflow, privacy, and long-term reliability measures as systems progress toward sustained deployment.}
\label{fig:benchmark_timeline}
\vspace{-8pt}
\end{figure}

This section proposes a staged evaluation framework that connects interaction choices (Section \ref{Sec:Interaction}) and privacy controls (Section \ref{Sec:Privacy}) with measurable outcomes so that pilot studies can be compared over time and across sites. 
It is organised around four progressively complex deployment scenarios, plus a consolidated view of performance and reporting metrics: simulations and offline evaluations, lab-based trials, pilot studies, and long-term deployments. 
Fig.~\ref{fig:benchmark_timeline} provides a simple illustration of these stages, and clarifies which families of measures become most informative as systems progress toward sustained autonomy. 
Across all stages, standard machine learning metrics such as accuracy, sensitivity, or area under the curve remain necessary but insufficient in aged care HRI. To support consistent comparison across sites, teams should carry a core set of operational measures through the full ladder, including safety events and near misses, false alarm burden per resident per day, caregiver workload change, time to recover after failures, privacy friction such as mute, pause, or opt out actions, and acceptability traces such as minutes of use per day and participation rate over weeks. For social and companion robots, additional indicators include reports on acceptability, usability, discomfort or rejection (e.g., uncanny valley effects), and how changes in appearance, voice, or behaviour impact engagement~\cite{damiano2018anthropomorphism}.

%\DA{From Simon: I'd suggest break this into five subsections:
%- Simulations 
%  > Section 6.2
%- Lab-based trials
%  > Recrutuiment  and ethics with ...
%- Pilot Studies
%  > Health-style reporting ...
%- Long term deployments
%- Measuring performance and metrics
%  > What to capture beyond accuracy
%  > Performance metrics beyond accuracy
%  > Explainability and transparency
%  > Benchmarking toolkits and open standards
%The first little bit can serve as the intro. The rest get's moved around as noted above.}

% --- Figure ---

%-------
\subsection{Simulations and offline evaluations}
\begin{figure}[!t]
\centering
\includegraphics[width=0.7\linewidth]{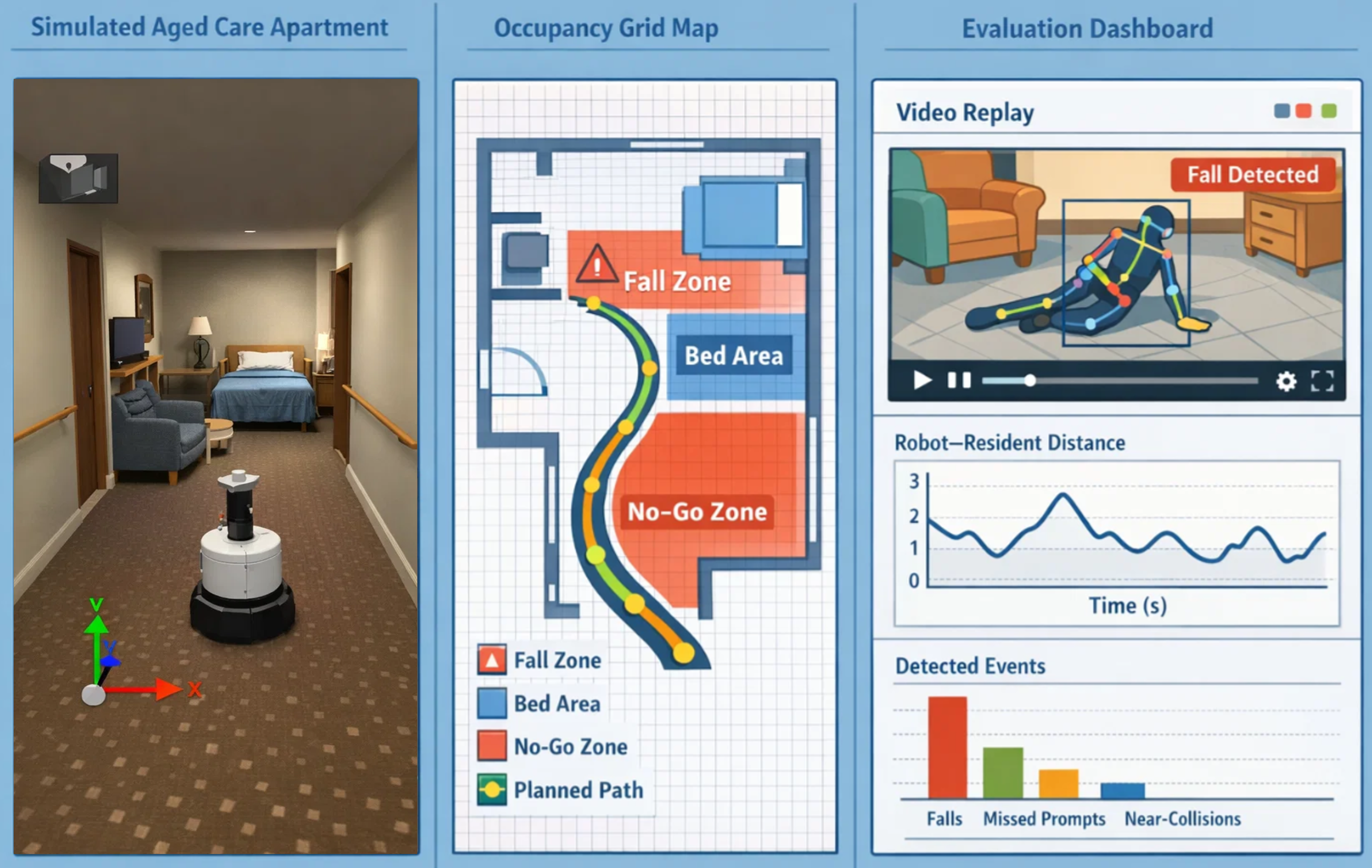}
\caption{
Simulation testbeds for assistive robots in aged care. 
Left: 3D simulation of a residential aged care apartment, with a mobile robot navigating around furniture, mobility aids, and narrow corridors. 
Middle: top-down occupancy grid and planned trajectory, with marked fall zones, bed areas, and no-go regions. 
Right: offline evaluation dashboard showing replay of robot–resident interactions alongside safety and performance metrics (distance to residents, detected events, and alert statistics) used to refine policies before real-world deployment.
}
\label{fig:simulation_collage}
\end{figure}

Simulation and dataset-driven evaluation provide a low-risk way to test feasibility and bound failure modes before any exposure to residents or staff. At this stage, teams can stress-test perception and planning under clutter, occlusion, and lighting variation, and establish baseline compute and power profiles that will later constrain on-device deployment. 
These ideas are often implemented in simulation testbeds that combine 3D care-home environments, mapping views, and offline evaluation dashboards for replay and analysis, as illustrated in Fig.~\ref{fig:simulation_collage}.

Results here should be treated as necessary evidence for progression to lab trials, not as proof of real-world reliability. 
Development and evaluation in an offline setting using simulation platforms or pre-collected datasets is a key first step for any new technology. Within the home care domain, a number of datasets and tools exist to support this stage. Key resources include: 

\begin{itemize}
    \vspace{-6pt}
    \item \textit{Ambient and smart-home datasets}. 
    The CASAS program provides large, longitudinal ambient datasets from real homes (motion, door, environmental sensors with some activity labels). Newer releases extend through 2024 and are useful for routine and change-point modelling, but have limited video or audio data for privacy reasons~\cite{cook2015activity,cook_2025_15708568}. These datasets are helpful for testing routine detection, anomaly detection, and basic workload estimates before moving to in-person trials.
    \vspace{-6pt}
    \item \textit{Falls and activity recognition datasets (vision)}. 
    Recent reviews catalogue vision-based fall datasets and methods; these reveal small sample sizes, staged data, and limited demographics. While datasets are useful for model pruning or ablation, they offer no proof of reliability. When using such data, one should expect domain shift in real deployments and plan for small, local validation sets with older adults~\cite{benkaci2024vision,inturi2025technical}.
    \vspace{-6pt}
    \item \textit{Embodied and simulation test beds}. 
    For navigation and manipulation in houses, Habitat (fast photorealistic rendering)~\cite{savva2019habitat}, AI2-THOR (interactive kitchen/living scenes)~\cite{kolve2017ai2}, iGibson and OmniGibson (interactive, physics-rich home scenes)~\cite{li2021igibson}, and BEHAVIOR-1K (everyday household activities with evaluation protocols)~\cite{li2023behavior} provide rich test beds. These tools are useful for stress tests and ablations before field trials, for example, checking how quickly a robot recovers from localisation loss or how navigation time changes with clutter. However, teams should still plan for the reality gap and validate behaviours with small staged pilots in real-world settings.
\end{itemize}

\noindent\textbf{Gaps remain:}
Privacy-labelled HRI datasets with per-frame tags are rare. Long-term multi-modal sequences with ground truth are scarce, especially when synchronised with caregiver actions, clinical outcomes, or workload. Datasets with older voices and atypical speech patterns are also limited, as are corpora that combine robot logs with ambient sensors and user-reported outcomes. These gaps motivate careful local data collection under the privacy safeguards discussed in Section~\ref{Sec:Privacy}.

%-------
\subsection{Lab-based trials}
Lab trials extend simulations into small-scale, highly controlled environments. The goal is to validate safety-critical interaction loops, sensor calibration, and consent practices with representative users before introducing full household or ward complexity. These trials should adopt proportionate, plain-language consent and capacity-aware processes, following national guidance for adults who may lack decisional capacity (e.g., Australia’s NHMRC National Statement~\cite{NHMRC_NationalStatement_2025}). 
These frameworks specify ongoing consent checks and data minimisation for audio or video, which aligns with the privacy by design approach in Section~\ref{Sec:Privacy}. Teams should also validate safety-critical interaction loops, such as emergency stop behaviour, conservative approach distances, and basic fault recovery drills, before exposure to full ward routines.

%-------
\subsection{Pilot studies}
Pilot studies evaluate feasibility in situ over weeks or months, and provide early signs of efficacy. 
In line with Fig.~\ref{fig:benchmark_timeline}, pilots should prioritise workflow fit, false alarm burden, early acceptability, and practical privacy controls that residents and staff can understand and reverse. 
When a system may influence care decisions, for example, through alerts, adherence reports, or behaviour change coaching, study protocols and reports should follow existing health-AI guidelines and protocols for transparency in algorithm updates, input handling, and human oversight (e.g., SPIRIT AI~\cite{rivera2020guidelines} and CONSORT AI~\cite{liu2020reporting}). For interventions that aim to change behaviour, brief descriptions following the TIDieR checklist help others reproduce content, dose, and facilitation. 
Cross-domain lessons from industrial HRI (e.g., near-miss recording, structured risk assessment) and smart home research (e.g., routines and change detection metrics) can inform pilot designs~\cite{lu2025impact,duarte2024benchmarking,lorenzini2023ergonomic,wu2023evaluating,yao2023reviewing}. 
However, consent and governance models present within industrial settings are often lighter than those expected in care homes, so any imported methods should be adapted with clinical and ethical oversight.

%-------
\subsection{Long-term deployments}
Long-term deployments reveal impact, reliability gaps, and maintenance burdens that short trials rarely expose, building on the earlier stages in Fig.~\ref{fig:benchmark_timeline}. From the outset, teams should plan to log errors explicitly, including docking misses, network drops, time drifts, and the time taken to recover from failure~\cite{hawes2017strands}. 
This stage is also where engagement stability, alert burden, and real caregiver workload effects can be assessed with enough duration to separate novelty from sustained value. Moreover, it provides the opportunity to conduct randomised-controlled trials~\cite{lu2025impact}, offering a rigorous evaluation of impact and efficacy beyond short-term pilots. 

%-------
\subsection{Measuring performance and metrics}
Evaluation should evolve without discarding core signals. Beyond the core measures introduced earlier, additional dimensions include economic considerations, performance beyond accuracy, privacy and security assurance, ethical and regulatory compliance, explainability and transparency, and emerging benchmarking toolkits and open standards.

\subsubsection{Economic considerations}
For costly automation systems affecting staff workload and service models, track the total cost of ownership. This should include installation time, staff training, replacement parts and consumables, cleaning and infection control, software updates, and support hours. Long-term studies (e.g., STRANDS) show that simple behaviours with reliable docking and easy staff customisation produce more sustained use than feature-rich systems that require frequent intervention~\cite{hawes2017strands}. Maintenance logs can be used to compute mean time between failure and mean time to repair, autonomous hours, the number of remote resets, and the number of support tickets per week.

\subsubsection{Performance metrics beyond accuracy}
Accuracy alone is insufficient. Modern benchmarks increasingly incorporate latency, energy consumption, robustness under occlusion, and adaptability to routine changes. Latency is critical for interactive systems, where sub-second response times often determine usability and safety. Energy efficiency matters for mobile robots and edge devices operating under power constraints, while robustness under occlusion and clutter ensures reliability in small, shared spaces. Adaptability metrics capture how systems maintain performance when furniture moves, lighting changes, or residents begin using new mobility aids. 
Fairness audits are critical to avoid demographic bias. Frameworks such as CLEAR (Cost, Latency, Efficacy, Assurance, Reliability) for enterprise AI~\cite{mehta2025beyond} illustrate how technical metrics can be tied to operational goals; an analogous approach can be used for care robots. 
As large language models are used for dialogue and planning become more common, pilots should also track conversational safety, hallucination-resistant escalation, and the quality of human handover when system confidence is low.

\subsubsection{Privacy and security benchmarks}
Privacy-preserving design is a cornerstone of ethical deployment and links directly to Section~\ref{Sec:Privacy}. Benchmarks can quantify the proportion of local versus cloud processing, the fraction of events logged as summaries rather than raw audio or video, and the typical retention period for different data types. Anonymisation and pseudonymisation methods should be checked with spot audits.
Security stress tests evaluate resilience against threats such as unauthorised access, replay attacks on networked sensors, or adversarial inputs that trigger false alerts. Guidance from security communities (e.g., OWASP \cite{bunzel2024bridging} and NIST \cite{ahmed2024cybersecurity}) emphasises layered defences, encryption at rest and in transit, and robust identity management for AI-enabled applications. For generative and perception systems, runtime safeguards such as input validation and anomaly detection help prevent leakage of sensitive information. These benchmarks should be aligned with international privacy frameworks (e.g., GDPR, HIPAA) and technical reports such as ETSI TR 104 225, which outline privacy aspects of AI/ML systems \cite{ahmed2024cybersecurity}.

\subsubsection{Ethical and regulatory compliance checks}
AI influencing care decisions must follow reporting standards like SPIRIT-AI \cite{rivera2020guidelines} and CONSORT-AI \cite{liu2020reporting}, ensuring transparency in algorithm description, input handling, and human-AI interaction. These extensions address AI-specific risks such as bias, drift, and error analysis, complementing traditional clinical trial standards and benchmarks for AI-enabled assistive technology \cite{Silvera-TawilAIFramework}. For reproducibility of behaviour change interventions, the TIDieR framework specifies reporting of content, dose, and facilitation. Compliance checks should also include privacy impact assessments and consent audit trails, particularly for vulnerable populations, to meet national and international ethical requirements. 

\subsubsection{Explainability and transparency}
Benchmarks assess explanation fidelity (alignment with actual model logic) \cite{yeh2019fidelity}, user comprehensibility, and compliance with regulatory principles like those outlined in NIST’s Four Principles of Explainable AI \cite{quinn2020recommended}. Tools such as SHAP \cite{lundberg2017unified} and LIME \cite{ribeiro2016should} provide quantitative measures of feature attribution, while domain-specific frameworks emphasise interpretability in safety-critical contexts like healthcare and robotics. Transparency audits should ensure explanations are resistant to manipulation and meaningful for end-users, balancing interpretability with operational performance.

\subsubsection{Benchmarking toolkits and open standards}
Frameworks are emerging to support reproducible and fair evaluation of AI systems. MLPerf \cite{mattson2020mlperf}, developed by MLCommons, offers benchmarks for inference and training across cloud and edge platforms, including latency and throughput tests relevant to real-time care robotics. However, gaps remain in HRI-specific metrics, particularly for social interaction quality, privacy compliance, and long-term autonomy. Future standards should integrate technical, ethical, and human-centric indicators to enable comprehensive evaluation of care robotics.

%-----------------------------------------------------------------------------------

%\clearpage
% --------------------------------------------------------------
\section{Open research challenges}
\label{Sec:open_research_challenges}
\begin{figure}[!b]
\centering
\includegraphics[width=0.8\linewidth]{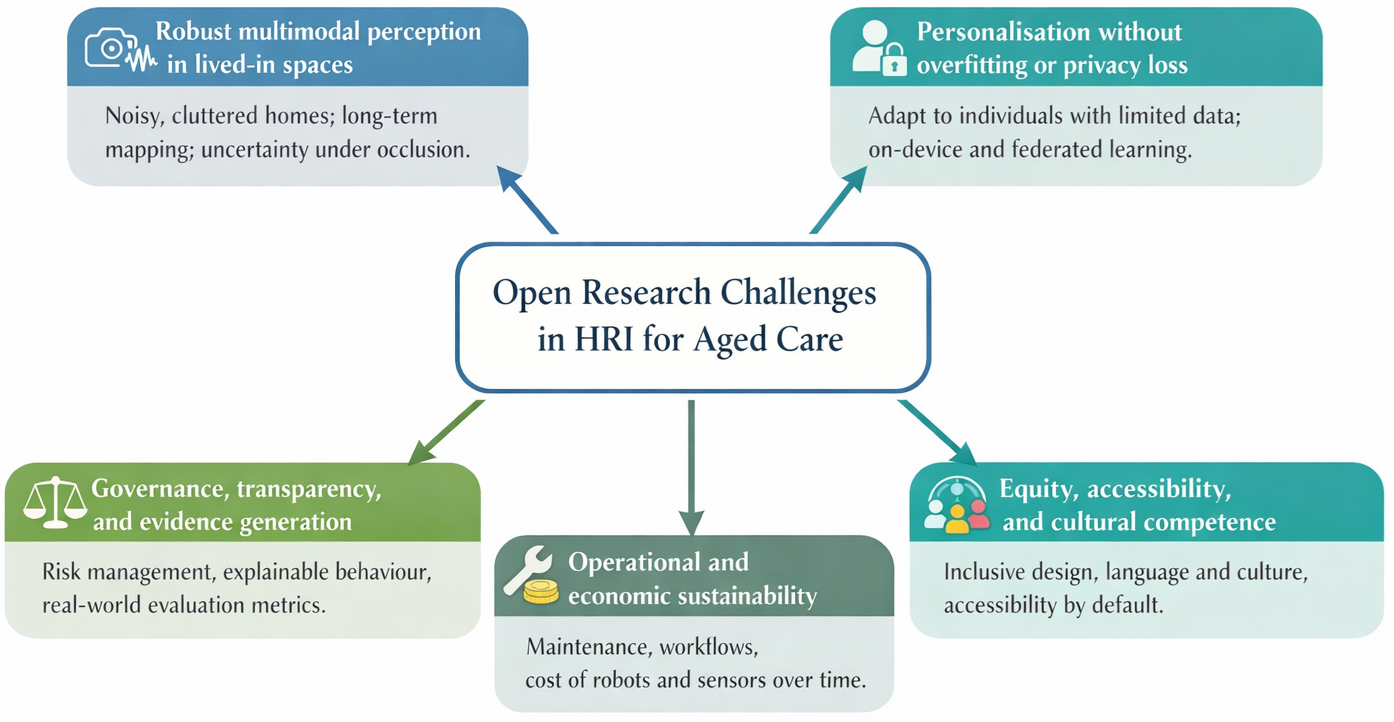}
\caption{
Schematic overview of five open research challenges in assistive human–robot interaction for private and residential aged care homes: robust multimodal perception, personalisation without privacy loss, governance and evidence, operational and economic sustainability, and equity and cultural competence.
}
\label{fig:open_challenges}
\end{figure}

Despite rapid progress in sensing, perception, and interaction design, deploying assistive robots and ambient intelligence in private and RAC homes remains a demanding frontier. These environments expose gaps in robustness, personalisation, safety, and governance that current systems only partially address. We highlight five open challenges that require sustained research attention, summarised in Fig.~\ref{fig:open_challenges}.

%-------------------------------
\subsection{Robust multimodal perception in lived‑in spaces}
Homes and RAC facilities remain difficult environments for perception. Lighting varies, furniture moves, and people use mobility aids that occlude bodies or alter gait. Radar, thermal, and depth sensors help, but fusing these signals reliably under uncertainty is still an open problem. Systems that perform well in controlled settings often struggle when views are partial, audio is noisy, or sensors are intermittently occluded. Radar, thermal, and depth sensors can mitigate some of these issues, but reliably fusing heterogeneous signals under uncertainty is still an open problem.

A central difficulty is multimodal fusion under uncertainty and occlusion. Domestic environments routinely produce partial views, motion blur, and conflicting signals from voice, vision, and ambient sensors. The unsolved challenge is ensuring stable behaviour when inputs disagree or drop out, and presenting uncertainty in ways that older adults and non‑technical staff can interpret quickly without increasing cognitive load or alarm fatigue. These issues connect directly to the interaction patterns in Section \ref{Sec:Interaction} and depend on the sensing and learning choices described in Section \ref{Sec:Technologies}.
    
Sustaining performance as conditions drift over days and weeks is another open problem. This links directly to the interaction patterns in Section~\ref{Sec:Interaction} and will draw on sensing and learning choices in Section~\ref{Sec:Technologies}. Long-running autonomy studies show that real performance depends on recovery after faults, docking reliability, and the ability to continue service under network loss \cite{hawes2017strands}. Yet the field lacks methods to quantify how robustness degrades over time and how those shifts translate into user‑visible impacts such as false alarms and missed assistance.

Generalisation across homes is equally difficult. Models trained in one set of environments often fail in another due to differences in layout, objects, accents, and routines. Robust domain‑shift handling and simple ways to communicate uncertainty to users and carers are essential for deployments that span both private homes and RAC facilities, where procurement expects repeatable performance despite local variation.

These challenges extend to navigation and manipulation as well. Domestic spaces are dense and narrow. Walkers, wheelchairs, cables, and pets create hazards. Dependable navigation and manipulation near people, with predictable approach distances and safe, reversible handovers in small rooms, remains challenging. Navigation in such spaces remains a limiting factor for many tasks that require physical help \cite{duan2024human}. 

Finally, many activities can be sensed without a robot, using vision and ambient systems for falls, routines, and unusual events. The open question is how to transfer these strengths into human-robot interaction without losing privacy and clarity. We need principled ways to determine when a robot adds value beyond a fixed sensor network, and when it does not, using measures that matter in real life \cite{benkaci2024vision}. This comparison should include governance burden, not only technical performance.

Together, these gaps point to the need for advances in multimodal fusion, online calibration, fault‑tolerant mapping, and uncertainty‑aware interaction, all operating without expert intervention and all understandable to older adults and non‑technical staff.

%-------------------------------
\subsection{Personalisation without overfitting or privacy loss}
There are open questions on how to personalise systems safely over time while respecting privacy and working with very limited data. Older adults differ in language, hearing, vision, cognition, routines, and preferences. Personalisation improves usability but raises privacy concerns and can lead to brittle models when data is scarce. Methods such as federated learning, continual adaptation, and uncertainty‑aware policy updates show promise, yet remain difficult to deploy safely in small, heterogeneous datasets typical of home environments \cite{casado2024towards}. 
Balancing adaptation with predictability---and doing so transparently---remains a central challenge.
    
This connects to the privacy by design in Section~\ref{Sec:Privacy} and to edge learning choices detailed in Section~\ref{Sec:Technologies}. Equally important is safe rollback and auditability, so that a system can revert to a stable baseline when an adaptation harms performance, safety, or user comfort.

%-------------------------------
\subsection{Governance, transparency, and evidence generation}
Regulation for AI‑enabled care technologies is evolving rapidly. Designers must demonstrate safety, privacy, and fairness while keeping systems understandable to residents, families, and staff. This creates open questions around explainability for non‑experts, auditable decision pipelines, and evaluation protocols that reflect real‑world risk. There is also a need for shared and unified benchmarks, reference architectures, and deployment playbooks that help teams generate regulation‑ready evidence without reinventing foundational components.

Evidence gaps are closely linked to the dataset and benchmark limitations summarised in Section~\ref{Sec:Benchmark}. Few studies report benefits from large‑scale, long‑term deployments across diverse sites, and many rely on proxy metrics such as engagement rather than clinical or operational outcomes. Health style reporting, such as SPIRIT AI and CONSORT AI, helps reviewers judge bias, drift, and the role of human oversight, but adoption in human-robot interaction remains uneven \cite{liu2020reporting,rivera2020guidelines}. Section~\ref{Sec:Benchmark} proposed a multi‑dimensional suite of metrics that could be used to evaluate performance. The open methodological challenge is to normalise these measures across heterogeneous sites so that results can be pooled or compared without losing context, while still respecting local privacy, workflow, and staffing constraints.

Dataset limitations further constrain governance and evidence generation. There are very few privacy-labelled HRI datasets, few long-horizon multi-modal recordings with ground truth, and limited speech data from older voices and atypical speech for automatic speech recognition. Public datasets that link robot logs, ambient sensors, and human outcomes are rare. Although simulation resources cover many domestic scenes, the reality gap remains, and there is a need for structured datasets and protocols that bridge from small-scale pilots to in-the-wild trials \cite{cook2015activity,cook_2025_15708568,savva2019habitat,li2021igibson,li2023behavior}. A further gap is the limited availability of datasets that document configuration changes, staff handovers, and consent state over time, even though these factors strongly shape real-world reliability and acceptability.

\subsection{Operational and economic sustainability}
Upfront costs attract attention, but the true cost of ownership includes installation, training, cleaning, consumables, software updates, and support. In practice, systems that are slightly less capable but much easier to maintain often outperform more advanced alternatives. Open questions include how to design systems for low support burden, how to predict maintenance from telemetry, and how to compare robots with non-embodied alternatives in honest cost-benefit terms. Lessons from long-running deployments suggest that simple behaviours with reliable docking and easy staff configuration sustain use better than complex features that require tinkering \cite{hawes2017strands}. A persistent research gap is the lack of standard reporting on maintenance time, remote‑support frequency, and failure categories in domestic trials, which limits realistic comparison across platforms and sites.

%-------------------------------
\subsection{Equity, accessibility and cultural competence}
Older adults are not a homogeneous group. Language, culture, income, and digital literacy shape expectations and comfort. Accessibility defaults such as large targets, announced state changes, and visible controls help, but the field lacks strong evidence on culturally competent interaction styles that remain transparent and do not mislead. The open task is to build libraries of content and behaviours that can be tailored by staff without technical help and that are evaluated for fairness and inclusion over months rather than days \cite{W3CWCAG22,slane2025bringing}. This requires multilingual and low‑literacy evaluation protocols that treat accessibility as a core performance dimension, and that assess whether adaptation and personalisation benefit all user groups equally rather than reinforcing existing disparities.

Equity also depends on access to hardware and infrastructure, not just on interface design. Robots, depth cameras and dense ambient sensor networks remain expensive capital items and typically require reliable Wi-Fi, sufficient power, secure networking and specialist technical support. Larger metropolitan providers or well-resourced pilots can often absorb these costs; whereas smaller organisations, rural and remote services and low-income households may only be able to deploy basic pendant alarms, tablets or consumer smart speakers. Without careful design, this risks concentrating advanced HRI capabilities in already advantaged sites and widening gaps in quality of care.

To mitigate this, extreme-environment HRI work should prioritise modular systems that can reuse existing infrastructure (for example, televisions, off-the-shelf tablets and home routers), operate gracefully under intermittent connectivity, and be serviced remotely rather than relying on onsite engineers. Procurement and funding models should explicitly budget for long-term maintenance, device refresh and staff training, so that accessibility to robots and sensing infrastructure does not depend solely on local champions or short-term grants.

%-----------------------------------------------------------------------------------

%\clearpage
% --------------------------------------------------------------
\section{Future research directions}

The previous section outlined open challenges that currently hinder the dependable use of HRI in RAC and private home environments. These challenges span robustness in lived-in spaces, personalisation with privacy, risk management, evidence beyond accuracy, and questions of equity and cultural fit, but they also point to a smaller set of practical research levers that can help to close multiple gaps at once. In this section, we focus on a smaller set of shared directions that can help the field move from promising prototypes to dependable practice. Rather than mapping one future direction to each open challenge, we describe a selection of cross-cutting levers that can support many projects: reference architectures, privacy-aware personalisation, evaluation ladders and benchmarks, operational tooling, and long-term co-design with stakeholders.

\subsection{Reference architectures for hybrid robot and ambient homes}
Many tasks can be met by ambient sensing, while others truly require a robot. A shared, modular architecture would help teams combine both approaches while keeping boundaries between sensing, decision-making, and data sharing explicit. At minimum, such an architecture should define how local perception and intent run on-device, how events are filtered into minimal summaries for a local care gateway that aggregates devices within a private home or a facility unit, what data are retained locally versus forwarded, and when to escalate to clinicians through standards aligned with health data exchange, including FHIR resources such as \textit{Observation} or \textit{Detected Issue} \cite{pavao2023fast,li2024review}. Development could be supported by publishing a small set of event schemas covering common tasks such as falls, adherence, and welfare checks; and by defining latency, access, and retention targets that match the privacy by design principles from Section~\ref{Sec:Privacy}. Where systems interface with clinical services, these schemas should be documented in ways that support audit and safe handover, so that the same interaction and privacy assumptions can be traced from sensing to alerting to review~\cite{oaicAPPs2022}.

\subsection{Personalisation that stays private}
The field needs worked examples of long-horizon adaptation that run for months with very little labelled data, clear consent, and conservative safety envelopes. Promising directions include federated or split learning for voice and activity models, identity-disentangled representations for expression or engagement cues, and privacy-preserving training that never pools raw personal data \cite{casado2024towards,brauneck2023federated}. To make these approaches credible in care settings, future systems should publish explicit adaptation budgets, safe rollback mechanisms, and simple tests that detect when personalisation risks increase false alarms or reduce user comfort. Each method should include simple diagnostics that show when the system is uncertain, how behaviour was adapted, and how a resident or caregiver can pause, reset, or override the model. To remain consistent with the governance constraints outlined earlier in Sections ~\ref{Sec:Privacy} and ~\ref{Sec:open_research_challenges}, adaptation should be paired with clear role-based controls that distinguish resident, family, and staff permissions, and log when personalisation is enabled, paused, or reverted.

\subsection{Evaluation ladders that bridge the reality gap}
Research should develop shared evaluation ladders that move from simulation to lab checks to micro-pilots and then in-the-wild studies, with a small set of core measures carried across stages, as emphasised in Section~\ref{Sec:Benchmark}. For simulation, development can lean on domestic platforms such as Habitat, AI2-THOR, and OmniGibson to stress-test perception and policies \cite{savva2019habitat,li2021igibson}. For lab and early pilots, staged sessions that test occlusion, clutter, and audio noise, and verify privacy controls before broader deployment are required. For field studies, adopting health‑style reporting allows reviewers to judge bias, drift, the role of human oversight, and how missing data were handled \cite{liu2020reporting,rivera2020guidelines}. Recent long-horizon studies in care contexts also reinforce the value of pairing these reporting standards with practical operational measures, such as maintenance burden and staff-facing configuration logs, to support comparability across sites \cite{otaka2024longitudinal}.

\subsection{Co-design with caregivers and residents that lasts beyond the pilot}
Early co-design improves fit, but long-term fit depends on who can change content, schedules, and thresholds after the researchers leave. Future studies should provide simple tools for staff and family to configure systems, use planned reviews with caregivers every few weeks, and report what settings were changed in practice and how those changes affected safety events, false alarm burden, and engagement. Evidence from recent care home deployments suggests that staff perceptions of workload, trust, and accountability evolve over time, so co-design should be treated as a continuing operational capability and not only a pre-deployment activity \cite{ruan2025scoping,otaka2024longitudinal}.

\subsection{Summary}
Together, these future directions suggest a practical roadmap from prototypes to regulation-ready evidence. In the near term, teams can publish and adopt shared event schemas, on-device processing patterns, and minimal data flows that make systems easier to integrate and audit. In the medium term, privacy-aware personalisation should be demonstrated in multi-month deployments with clear uncertainty signals and low-burden override tools for residents and staff. In the longer term, multi-site studies that follow SPIRIT AI and CONSORT AI can link technical improvements to measurable outcomes such as reduced false alarms, lower caregiver workload, and safer recovery after faults. This pathway builds directly on the interaction, technology, privacy, and benchmarking foundations established earlier in the chapter, and keeps the emphasis on systems that remain useful, respectful, and sustainable in everyday care.
    
%-----------------------------------------------------------------------------------

%%%%%%%%%%%%%
%\clearpage
\section{Conclusion}

This chapter examined how robots and ambient intelligence can support older adults in RAC and private homes, focusing on the technical, human, and organisational constraints that shape real‑world deployment. We argued that these environments constitute extreme HRI settings because they are cluttered, dynamic, have low support, and are governed by strong expectations of safety, privacy, and dignity. Designing for such contexts requires more than accurate perception or engaging interaction; it demands systems that remain robust, transparent, and respectful throughout everyday use.
Three themes emerged across the chapter. First, robust perception and interaction in lived‑in spaces are essential, given occlusion, mobility aids, shifting layouts, and diverse sensory and cognitive needs. Second, privacy, trust, and acceptability must be built into system architecture and interaction design from the outset. Third, evaluation must extend beyond accuracy, incorporating metrics that reflect the realities of care, such as burden, recovery, maintainability, and long‑term reliability.
We also emphasised that dependable deployment depends on hybrid thinking, including operational fit and governance (not only technical capability). Hybrid robot‑and‑ambient architectures, privacy‑aware personalisation, evaluation ladders, and long‑term co‑design can help bridge the gap between prototypes and sustained use, while supporting regulation‑ready evidence and interoperability with existing care systems.
Evaluation must also reflect real-life outcomes. Practical study ladders should link simulation and lab tests to in-the-wild pilots and long-term deployments, with common measures such as false alarms per resident per day, caregiver workload, recovery after faults, and sustained engagement. These metrics help teams compare systems across sites and ensure that improvements translate into meaningful benefits for residents and staff. 
Ultimately, the goal is to create assistive technologies that support ageing with dignity, safety, and independence. By grounding innovation in the realities of home and care environments, and by treating privacy, trust, and usability as core engineering requirements, the community can move toward systems that are not only intelligent but genuinely welcome in the places people live.

%\clearpage
%\input{sections/0_Literature_list}
%-----------------------------------------------------------------------------------

%------------
\bibliographystyle{Harvard}

{\footnotesize
\bibliography{ref}}

\end{document}